\documentclass[aps,twocolumn,superscriptaddress,showpacs]{revtex4-1}

\usepackage{times}
\usepackage{amsmath}
\usepackage{amsfonts}
\usepackage{amssymb}
\usepackage{graphicx}
\usepackage{hyperref}
\usepackage{multirow}
\usepackage{hyperref}

\newcommand{\mnras}{Mon. Not. RAS}
\newcommand{\araa}{Ann. Rev. Astron. Astrophys.}
\newcommand{\ang}{\mathcal{L}_z}
\newcommand{\en}{\mathcal{E}}
\newcommand{\Q}{\mathcal{Q}}

\begin{document}

%\preprint{}

\title{Isofrequency pairing of geodesic orbits in Kerr geometry}


\author{Niels Warburton}
\affiliation{School of Mathematical Sciences and Complex \& Adaptive Systems Laboratory, University College Dublin, Belfield, Dublin 4, Ireland}
\affiliation{School of Mathematics, University of Southampton, Southampton, SO17 1BJ, United Kingdom}
\author{Leor Barack}
\affiliation{School of Mathematics, University of Southampton, Southampton, SO17 1BJ, United Kingdom}
\author{Norichika Sago}
\affiliation{ Faculty of Arts and Science, Kyushu University, Fukuoka 819-0395, Japan}
%\thanks{}


\date{\today}

\begin{abstract}
	Bound geodesic orbits around a Kerr black hole can be parametrized by three constants of the motion: the (specific) orbital energy, angular momentum and Carter constant. Generically, each orbit also has associated with it three frequencies, related to the radial, longitudinal and (mean) azimuthal motions. Here we note the curious fact that these two ways of characterizing bound geodesics are not in a one-to-one correspondence. While the former uniquely specifies an orbit up to initial conditions, the latter does not: there is a (strong-field) region of the parameter space in which pairs of physically distinct orbits can have the same three frequencies. In each such isofrequency pair the two orbits exhibit the same rate of periastron precession and the same rate of Lense-Thirring precession of the orbital plane, and (in a certain sense) they remain ``synchronized'' in phase.
\end{abstract}

%\pacs{{\bf [check]}}
%\keywords{}

\maketitle


\section{Introduction} 

The motion of test bodies in the Kerr metric of a rotating black hole has been studied for almost half a century (see, e.g., \cite{Carter,Bardeen-Press-Teukolsky,Wilkins,Chandra,Bicak_etal}). Much of the more recent work is motivated by the need to understand radiative inspirals into a Kerr black hole as sources of gravitational waves for future detector experiments. Examples of recent work include an action-angle formalism \cite{Schmidt}, a frequency-domain method for computing functionals of the orbit (such as the gravitational perturbation from an orbiting test particle) \cite{Drasco-Hughes}, a system for classifying Kerr orbits \cite{Levin-Periz-Giz:BH_periodic_table,Grossman-Levin-Periz-Giz}, and an analytic method for solving the geodesic equations of motion \cite{Fujita-Hikida}. 

Timelike geodesics of the Kerr geometry are completely integrable. They admit three nontrivial constants of motion (``first integrals''), each associated with a Killing field of the Kerr background: the time-translation and rotational Killing vectors give rise to conserved (specific) orbital energy $\en$ and azimuthal angular momentum $\ang$, and the second-rank Killing tensor discovered by Carter \cite{Carter} gives rise to what is known as the Carter constant, $\Q$. Up to initial conditions, these three constants of motion uniquely label all timelike geodesics of the Kerr geometry.  

This paper is concerned with the family of {\em bound} geodesic orbits. Each bound orbit is confined to the interior of a compact spatial torus given by $r_{\rm p}\leq r\leq r_{\rm a}$ and  $\theta_{\rm min}\leq \theta\leq \pi-\theta_{\rm min}$, where hereafter ${t,r,\theta,\varphi}$ are Boyer-Lindquist (BL) coordinates, $r=r_{\rm p},r_{\rm a}$ are two radial turning points (``periastron'' and ``apastron'', respectively), and $\theta=\theta_{\rm min},\pi-\theta_{\rm min}$ are two longitudinal turning points. Generically, the motion is ergodic, in the sense that a generic orbit will pass arbitrarily close to any point on the torus within a finite time $t$ (exceptional are ``resonant'' orbits, mentioned briefly below). The triplet $\{r_{\rm p},r_{\rm a},\theta_{\rm min}\}$ provides an alternative parametrization of bound geodesics, which is in a one-to-one correspondence with that of $\{\en, \ang, \Q\}$ \footnote{That the mapping $\{r_{\rm p},r_{\rm a},\theta_{\rm min}\}\leftrightarrow\{\en, \ang, \Q\}$ is one-to-one can be establishing in the following way. We first note that Schmidt \cite{Schmidt} provides formula for $\{\en, \ang, \Q\}$ in terms of $(p,e,\theta_\text{min})$, and that there is a bijection between $(p,e)\leftrightarrow\{r_{\rm p},r_{\rm a}\}$ (straightforward to see from Eqs.~\eqref{pe} and their inverse). Furthermore Eqs.~\eqref{EOMr} and \eqref{EOMq} imply that $r_1 \equiv r_a$, $r_2 \equiv r_p$ and $z_-$ (and hence $\theta_\text{min}$) are given uniquely in terms of $\{\en,\ang,\Q\}$. The existence of these relations asserts that the original mapping is one-to-one.}.

Generically, bound orbits are triperiodic, with three frequencies $\Omega_r$, $\Omega_{\theta}$ and $\Omega_{\varphi}$ associated with the motions in the radial, longitudinal and azimuthal directions, respectively. Of these, $\Omega_r$ and $\Omega_{\theta}$ are ``libration''-type frequencies, defined from the (average) radial and longitudinal periods, while $\Omega_{\varphi}$ is a ``rotation''-type frequency, describing the average rate at which the BL azimuthal phase $\varphi$ accumulates in time. We define the above frequencies with respect to BL time $t$; this is useful for many purposes, because $t$ is also the proper time of an asymptotically far static observer (e.g., a gravitational-wave detector).
 It is important to note that, in general, the orbital radius $r$ and polar angle $\theta$ of a given orbit are {\em not} (separately) periodic functions of $t$: the $t$-interval between successive periastron passages is not constant, and  the $t$-interval between successive $\theta=\theta_{\rm min}$ passages is not constant either. There is a choice of a time variable (the so-called ``Mino time''---see Sec.~\ref{sec:Kerr_freqs} below) in terms of which the radial and longitudinal motions completely separate and become precisely periodic. However, in terms of BL time $t$, the orbital periodicity can generally only be defined through an infinite time average (or, equivalently, through an average over the orbital torus \citep{Grossman-Levin-Periz-Giz}). We shall define the BL-time frequencies more precisely below, following Schmidt \cite{Schmidt} and Drasco and Hughes \cite{Drasco-Hughes}

The above general description simplifies in several special cases. If the ratio $\Omega_r/\Omega_{\theta}$ is a rational number (``resonant orbits''), then the trajectory traced by the orbit in the $r$--$\theta$ plane is closed (with a finite $t$-period), and the ergodicity property is lost. If the orbit is equatorial ($\theta={\rm const}=\pi/2$), then $\Omega_\theta$ loses its meaning, and the orbit becomes biperiodic with frequencies $\Omega_r$ and $\Omega_\varphi$; in this case the radial motion is strictly periodic, with a radial period $2\pi/\Omega_r$. Similarly, if the orbit is circular ($r_{\rm p}=r_{\rm a}$), then $\Omega_{r}$ loses its meaning, the orbit becomes biperiodic with frequencies $\Omega_\theta$ and $\Omega_\varphi$, and the longitudinal motion is strictly periodic with period $2\pi/\Omega_\theta$. (Orbits that are both equatorial and circular are singly periodic with frequency $\Omega_\varphi$.) Finally, in the special case of a Schwarzschild black hole, one can always set up the BL system so that the orbit is equatorial and biperiodic with frequencies $\Omega_r$ and $\Omega_\varphi$.

%In the case of equatorial (possibly eccentric) or circular (possibly inclined) orbits the radial and zenithal motion is strictly periodic with respect to coordinate time and it is straightforward to define $\Omega_r$ and $\Omega_\theta$. For generic orbits, that are both inclined and eccentric, the radial and zenithal motion is in general quasi-periodic. This quasi-periodicity is what gives rise to the ergodic nature of generic bound geodesic orbits in Kerr spacetime. So long as the ratio $\Omega_r/\Omega_\theta$ is not a rational number the test body will, given enough time, pass through all points on the orbital torus \cite{Drasco-Flanagan-Hughes}. Despite the coupled nature of the radial and zenithal motion for generic orbits, Schmidt was recently able to show that well defined orbital frequencies can still be obtained \cite{Schmidt}. This discovery has been particular useful as it allows for the expansion of various dynamical quantities into Fourier series \cite{Drasco-Hughes} which have then be used to study gravitational wave snapshots from extreme-mass-ratio inspirals \cite{Drasco-Hughes-2006}.

The purpose of this article is to challenge the commonly held notion (see, e.g., \cite{Drasco-2009}) that the trio of frequencies $\{\Omega_\varphi,\Omega_r,\Omega_\theta\}$ provides a good parametrization of generic bound geodesics in Kerr, i.e., one which is in a one-to-one correspondence with $\{\en, \ang, \Q\}$ or $\{r_{\rm p},r_{\rm a},\theta_{\rm min}\}$. We show that this is not the case: there are infinitely many pairs of ``isofrequency'' orbits, which are physically distinct (i.e., have different $\{\en, \ang, \Q\}$ values) and yet they share the same values of $\{\Omega_\varphi,\Omega_r,\Omega_\theta\}$. This point was already made briefly by two of us in Appendix A of \cite{Barack-Sago-2011} in reference to a Schwarzschild black hole (where orbits are biperiodic, and two isofrequency orbits share the same values of $\Omega_\varphi$ and $\Omega_r$). Here we first revisit the Schwarzschild problem to provide a further illumination of this phenomenon, and then extend the analysis to the Kerr case, showing that isofrequency pairing occurs even among triperiodic orbits. 

We shall on occasion refer to a pair of isofrequency orbits as ``synchronous'', because the phases of such orbits remain synchronized in an average sense. For example, two equatorial isofrequency orbits that pass through their periastra simultaneously at $\varphi=0$ will reach their next periastra at the same time and with the same azimuthal phase; they will have experienced an identical amount of periastron advance. Although such orbits go ``in and out of phase'' between periastron passages, their phase remains synchronized ``on average''. We will present some graphics to illustrate this behavior. 

Throughout this article we use geometric units such that the gravitational constant and the speed of light are both equal to unity. We denote the black hole's mass and spin by $M$ and $aM$, respectively. We use an over-tilde to denote adimensionalization using $M$; for example, $\tilde\Omega_{\varphi}:=M\Omega_{\varphi}$ and $\tilde a:=a/M$. We adopt a convention whereby $a>0$ and $a<0$ correspond to prograde and retrograde orbits, respectively, with $\ang$ always positive. We use the term ``orbit'' synonymously with ``timelike geodesic orbit''. In Sec.\ \ref{Sec:Schwarzschild} we consider (biperiodic) synchronous orbits in Schwarzschild geometry ($a=0$). We delineate the region in the parameter space where such orbits occur, and also provide an intuitive explanation as to why isofrequency pairing must occur. In Sec.\ \ref{Sec:Kerr} we generalize our discussion to the Kerr case, where we consider first equatorial orbits and then generic, triperiodic orbits.

\section{Isofrequency orbits in Schwarzschild geometry}\label{Sec:Schwarzschild}

\subsection{Orbital frequencies and separatrix}

The radial motion of geodesic test particles in the equatorial plane of a Schwarzschild black hole satisfies
\begin{equation}\label{EOM}
\dot{r}^2={\en}^2-V,\quad\quad
V(r;{\ang}):=\left(1-\frac{2M}{r}\right)\left(1+\frac{\ang^2}{r^2}\right),
\end{equation}
where a dot denotes differentiation with respect to proper-time, and $V(r;\ang)$ is an effective potential for the radial motion. Bound orbits exist for $\ang >2\sqrt{3}M$ with $\frac{2\sqrt{2}}{3}<\mathcal{E}<1$. For each $\{\en,\ang\}$ in this range, $\dot{r}^2(r)$ has three real roots, and motion is allowed between the second largest and largest of these, which we label $r_{\rm p}$ and $r_{\rm a}$, respectively. A convenient alternative parametrization of bound orbits is provided by the pair of values $\{p,e\}$ defined through
\begin{equation}\label{pe}
M p := \frac{2r_{\rm p} r_{\rm a}}{r_{\rm a}+r_{\rm p}}, 
\quad\quad
e := \frac{r_{\rm a}-r_{\rm p} }{r_{\rm a}+r_{\rm p}},
\end{equation}
which are relativistic generalizations of semi-latus rectum and eccentricity, respectively \cite{Darwin-1961}. This parametrization is in a one-to-one correspondence with that of $\{\en,\ang \}$. Explicitly,
\begin{eqnarray}
	\en^2 = \frac{(p-2-2e)(p-2+2e)}{p(p-3-e^2)},\quad \ang^2=\frac{p^2M^2}{p-3-e^2}, \label{eq:Schwarzschild_Energy_Ang_Mom}
\end{eqnarray}
which can be inverted (for real $p,e$) to give unique expressions for $p(\en,\ang)$ and $e(\en,\ang)$. In the $(p,e)$ space, bound orbits span the range $0\le e<1$ with $p\ge p_s(e):=6+2e$. The boundary $p_s(e)$ (``separatrix'') separates between stable and unstable orbits in the $(p,e)$ space \cite{Cutler-Kennefick-Poisson}. The $(p,e)=(6,0)$ terminus of the separatrix curve is known as the innermost stable circular orbit (ISCO). The existence of a separatrix is one of the salient features of motion in black hole spacetimes, and it marks a major qualitative departure from Newtonian dynamics. As we shall see, the occurrence of isofrequency pairing of orbits is intimately related to the existence of a separatrix.

The function $r(t)$ is periodic with ($t$-)period $T_r$. Following Darwin \cite{Darwin-1961}, it is convenient to introduce the ``relativistic anomaly'' parameter $\chi$, which is related to $t$ via 
\begin{equation}
%	\frac{d\varphi}{d\chi} 	&=& \sqrt{\frac{p}{p-6-2e\cos\chi}}\,,		\label{eq:dphi_dchi}	\\
	\frac{dt}{d\chi} 		= \frac{Mp^2[(p-2)^2-4e^2]^{1/2}(p-6-2e\cos\chi)^{-1/2}}{(p-2-2e\cos\chi)(1+e\cos\chi)^2}\, ,\hskip3mm	\label{eq:dt_dchi}
\end{equation}
and in terms of which the radial motion is given simply by $r(\chi)=Mp/(1+e\cos\chi)$ (taking $\chi=0$ at a periastron passage).
The radial period can then be computed via 
\begin{equation}
T_r 		= \int^{2\pi}_0 \frac{dt}{d\chi}\,d\chi\,, \label{eq:T_r}
\end{equation}
with associated radial frequency 
\begin{equation}
	\Omega_r := \frac{2\pi}{T_r} \, .		\label{eq:Schwarzschild_r_frequency}
\end{equation}

The {\it azimuthal} frequency of the orbit is defined as the average of $d\varphi/dt$ (with respect to $t$) over a complete radial period:
\begin{equation}
\Omega_\varphi := \frac{1}{T_r}\int_0^{T_r}\frac{d\varphi}{dt}dt=\frac{\Delta\varphi}{T_r}\, ,	\label{eq:Schwarzschild_phi_frequency}
\end{equation}
where $\Delta\varphi$ is the azimuthal phase accumulated over time interval $T_r$. The latter can be computed via
\begin{eqnarray}
	\Delta\varphi 	&=& \int^{2\pi}_0 \frac{d\varphi}{d\chi}\,d\chi = 
\int_{0}^{2\pi}\frac {\sqrt{p}}{\sqrt{p-6-2e\cos\chi}}\, d\chi \nonumber\\
&=&
	4\sqrt{ \frac{p}{\epsilon}}\, K\left(-\frac{4e}{\epsilon}\right)\,, \label{eq:DeltaPhi}		
\end{eqnarray}
%where the integrand is given by
%\begin{equation}
%	\frac{d\varphi}{d\chi} 	= \sqrt{\frac{p}{p-6-2e\cos\chi}}\, ,	\label{eq:dphi_dchi}	\\
%\end{equation}
where $\epsilon := p-p_s(e)$ and $K(x):=\int_{0}^{\pi/2}d\theta(1-x\sin^2\theta)^{-1/2}$ is the complete elliptic integral of the first kind.
%\footnote{Throughout this work our definition of the elliptic integrals coincides with that of Abramowitz and Stegun \cite{Abramowitz-Stegun}.}.

At the separatrix limit, $\epsilon\to 0^{+}$, both $\Delta\varphi$ and $T_r$ diverge at a similar rate [see Eqs.\ (\ref{DeltaphiAsy}) and (\ref{TrAsy}) below], so that $\Omega_r\to 0$ while $\Omega_{\varphi}$ attains a finite value [$=(M/r_{\rm p}^3)^{1/2}$, corresponding to the frequency of the unstable circular orbit of radius $r_{\rm p}=p/(1+e)$]. This gives rise to the well known ``zoom-whirl'' behavior \cite{Glampedakis-Kennefick}: orbits with $\epsilon\ll 1$ can ``whirl'' around the black hole many times near the periastron before ``zooming'' back out towards the apastron.

\subsection{Isofrequency orbits}

\begin{figure}
	\includegraphics{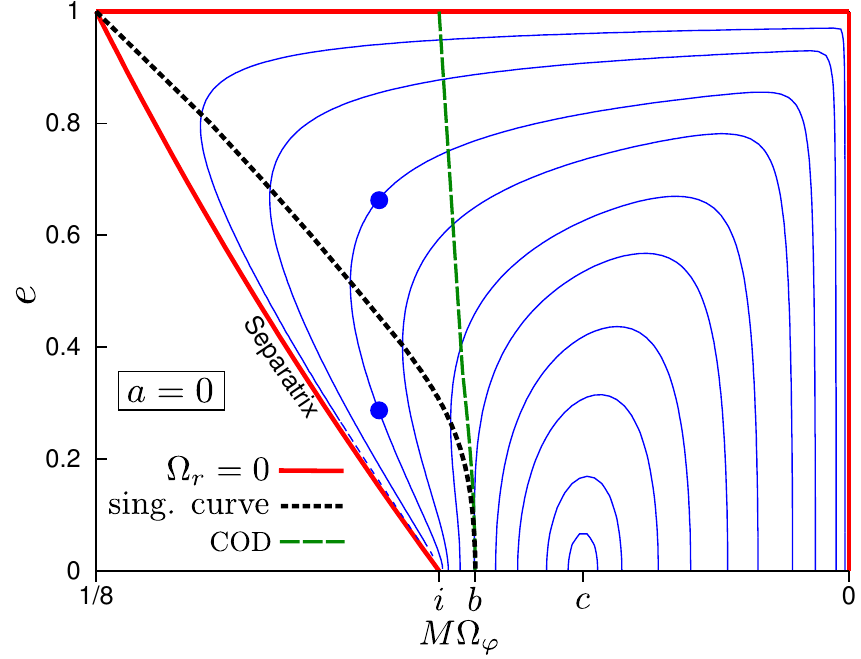}
	\caption{The $(\Omega_\varphi,e)$ parameter space for bound geodesic orbits in Schwarzschild geometry. Bound orbits are confined to the region right of the curve marked {\it separatrix}. Thin (blue) curves are contour lines of constant $\Omega_r$. The marginal contour line $\Omega_r = 0$ is shown as a thick (red) line. $\Omega_r$ takes its greatest value at the point marked $c$, representing a (slightly perturbed) circular orbit of radius $r_c=8M$. The dotted (black) line shows the curve along which the Jacobian matrix of the transformation $(p,e) \leftrightarrow (\Omega_r,\Omega_\varphi)$ becomes singular. The singular curve intersects the $e=0$ axis at $b$, corresponding to a circular orbit of radius $r_b = (39+\sqrt{145})M/8\simeq 6.3802M$. Any vertical ($\Omega_\varphi=\text{const}$) line left of $b$ intersects some $\Omega_r=\text{const}$ contours {\em twice}. Each pair of intersections identifies a pair of isofrequency orbits; a sample pair is marked in the plot. Each and every orbit between the separatrix and the singular curve has an isofrequency dual between the singular curve and the dashed (green) curve marked COD (for {\it circular-orbit duals}). The COD is the locus of all orbits dual to circular orbits of radius $r$ with $r_i=6M<r<r_b$.
%It can now be seen that pairs of synchronous orbits must exist as a contour of $\Omega_\varphi = \text{const} > b = (M/r_b)^{3/2}$, where $r_b = (39+\sqrt{145})/8M$, will intersect twice contours of $\Omega_r$ with $0<\Omega_r<[(r_b-6M)M)/r_b^4]^{1/2}$. 
%The (green) dotted line shows the circular orbit dual (COD) orbits with $\Omega_\varphi$ between $a=1/(6\sqrt{6})$ and $b$. The COD line also delineates the edge of the dual space. i.e., orbits to the right of this line can have no synchronous orbit dual, whereas orbits to the left of COD and the right of the $J=0$ line have a dual to the left of the $J=0$ line. We also mark on the point $c=8^{-3/2}$ where $\Omega_r$ takes its greatest value along $e=0$ (which is also the greatest value it attains anywhere within the parameter space).
	}\label{fig:Schwarzschild_Omega_phi_e}
\end{figure}

As pointed out in Ref.\ \cite{Barack-Sago-2011}, the Jacobian matrix of the transformation $(p,e) \to (\Omega_r,\Omega_\varphi)$ turns out to be singular along a certain curve in the parameter space, well {\em outside} the separatrix. This indicates that the transformation is not bijective. 
%It is a straightforward task to show that, for bound orbits, the parametrization transformation  $(\mathcal{E},\mathcal{L}) \leftrightarrow (p,e)$ is one-to-one. On the other hand the Jacobian, $J$, of the transformation $(p,e) \leftrightarrow (\Omega_r,\Omega_\varphi)$ is singular along a particular curve in the $(p,e)$ parameter space. The implication therefore is that it is possible to find two orbits, with different $(p,e)$, which share the same orbital frequencies. 
To see this most clearly it is instructive to move to a new orbital parametrization given by the pair $(\Omega_\varphi,e)$. This reparametrization is admissible because (i) as argued above, the original parametrization $(p,e)$ is a good one, and (ii) as can be easily checked, $\Omega_\varphi$ is a monotonically decreasing function of $p$ for any fixed $e$. Our argument now follows from examining the structure of the $\Omega_r = \text{const}$ contour lines in the $(\Omega_\varphi,e)$ plane, as shown in Figure \ref{fig:Schwarzschild_Omega_phi_e}. The key feature here is that some $\Omega_r = \text{const}$ contours have vertical tangents (the locus of which is shown by the dashed black line in the figure). Each of these contour lines is intersected {\em twice} by vertical lines just right of the vertical tangent. But 
vertical lines are also $\Omega_\varphi = \text{const}$ contours, and so the two intersections mark a pair of isofrequency orbits. (Any two such isofrequency orbits are clearly physically distinct: they have different eccentricities.) 

\begin{figure}
	\includegraphics[width=85mm]{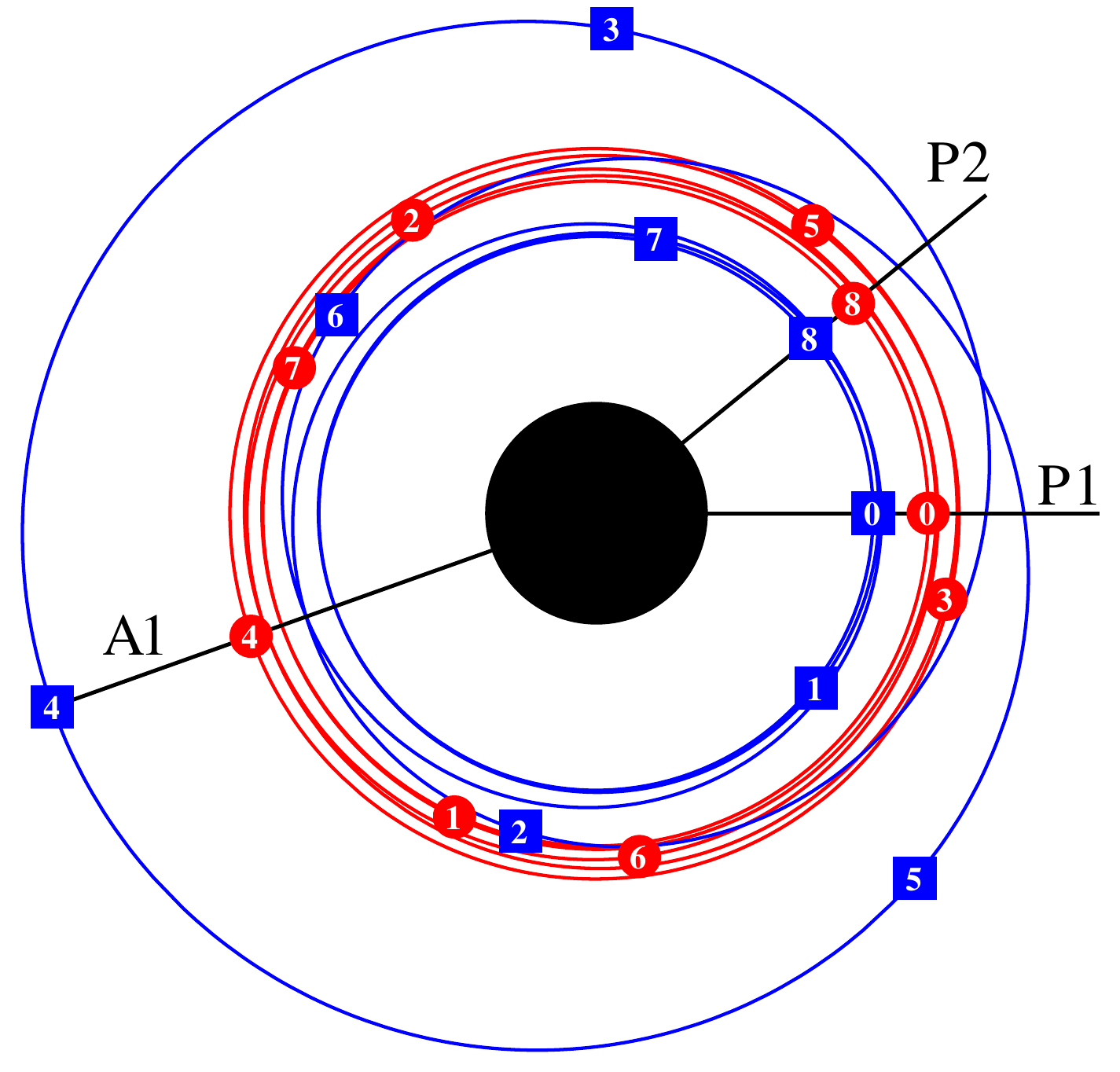}
	\caption{Orbital trajectories in the equatorial plane of a Schwarzschild black hole, for a sample pair of isofrequency orbits. The motion is anticlockwise, and the black hole is drawn to scale. Orbit 1 (red, round markers) has parameters $(p_1,e_1)=(6.255,0.05)$, and orbit 2 (blue, square markers) has parameters $(p_2,e_2)\simeq$ $(6.718788076, 0.3522488173)$. Both share the same orbital frequencies, $(\tilde\Omega_r, \tilde\Omega_\varphi) \simeq (0.01257801, 0.06426083)$. The orbital period of both orbits is $T_r\simeq499.535318M$ and each accumulates $\Delta\varphi\simeq 32.100669$ radians during that period. Both orbits start at their periastron marker `0' along the radial line $P1$. Each successive marker shows the orbital phase after a time period of $n\times T_r/8$, where $n$ is the marker number. At $T_r/2$ (marker 4) both orbits are synchronized again at their apastra along the line $A1$. When each test body has completed one orbit (marker 8) they are again synchronized at their periastra along the line $P2$. Both orbits have precessed by the same amount over their common radial period.}\label{fig:Schwarzschild_sample_pair}
\end{figure}

\begin{figure}
	\includegraphics[width=85mm]{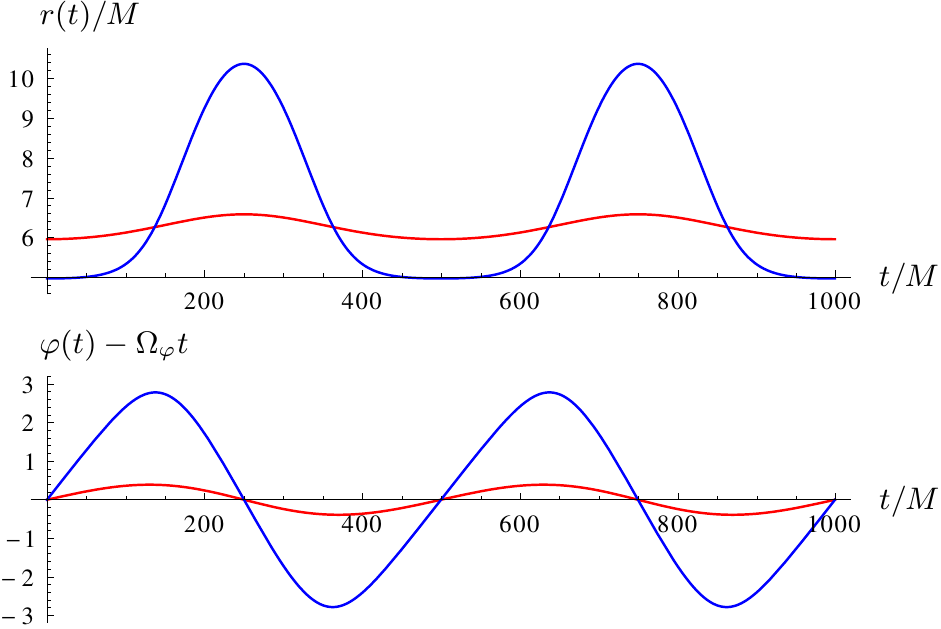}
	\caption{Evolution of $r(t)$ and $\varphi(t)-\Omega_\varphi t$ for the isofrequency pair shown in Fig.~\ref{fig:Schwarzschild_sample_pair}. Both radial and azimuthal motions are ``phase-synchronized'' on average. }\label{fig:Schwarzschild_example_r_phi}
\end{figure}

In Fig.~\ref{fig:Schwarzschild_sample_pair} we show, superimposed, the orbital trajectories of a sample pair of isofrequency orbits of rather different eccentricities. The radial and azimuthal motions of these two orbits are plotted in Fig.~\ref{fig:Schwarzschild_example_r_phi}. Since the rate of relativistic periastron advance depends only on the frequency ratio $\Omega_\varphi/\Omega_r$, two isofrequency orbits will exhibit the same rate of advance. This means that their phase remains ``synchronized'' on average, a behavior illustrated in the figures. 

Before giving a more detailed analysis, let us remark on the practicalities of producing the contour map of Fig.\ \ref{fig:Schwarzschild_Omega_phi_e}. The relation $\Omega_r(\Omega_\varphi,e)$ is not known analytically, so we resort to a numerical calculation: First, for a given $e$, we numerically invert the relation $\Omega_\varphi(p,e)$ [Eq.\ (\ref{eq:Schwarzschild_phi_frequency})] to find $p(\Omega_\varphi,e)$. 
Then we use Eq.~\eqref{eq:Schwarzschild_r_frequency}  to obtain $\Omega_r(p(\Omega_\varphi,e),e)$. Much of the interesting portion of the parameter space for our purpose lies very near the separatrix, where it becomes numerically challenging to evaluate the divergent quantities $T_r,\Delta\varphi$ and their ratio in Eq.\ (\ref{eq:Schwarzschild_phi_frequency}). In this problematic domain we instead use the near-separatrix analytic expansions \cite{Cutler-Kennefick-Poisson}
\begin{eqnarray} \label{DeltaphiAsy}
	\Delta\varphi 	&\approx& 	\sqrt{\frac{6+2e}{e}}\log\left(\frac{64e}{\epsilon}\right) + \mathcal{O}(\epsilon\log\epsilon)\,,		\\
	T_r 		&\approx& 	\frac{4M(3+e)^2}{\sqrt{e}(1+e)^{3/2}} \left[\log\left(\frac{64e}{\epsilon}\right)	\right.				\nonumber\\
				&+&			\left. \frac{\pi e (9+6e-7e^2)}{4(1-e^2)^{3/2}} + e I(e)\right] + \mathcal{O}(\epsilon\log\epsilon)\,. \label{TrAsy}
\end{eqnarray}
Here, the integral $I(e) := \int^\pi_0 (1+e \cos\chi)^{-2} D(\cos\chi)\,d\chi$, with
\begin{eqnarray}
	D(\cos\chi) &=& \frac{3+2e-e^2\cos^2\chi}{2+e(1-\cos\chi)}[2(1-\cos\chi)]^{1/2} 		\nonumber\\
				&-& 3+e-\frac{1}{4}(7e-3)(1+\cos\chi)\, ,	
\end{eqnarray}
is easily evaluated numerically. 
%By using these expansions near the separatrix and Eqs.~\eqref{eq:DeltaPhi} and \eqref{eq:T_r} otherwise it is possible to numerically calculate $\Omega_r$ for all points in the $(\Omega_\varphi,e)$ parameter space. The results are shown in Fig.~\ref{fig:Schwarzschild_Omega_phi_e} and clearly demonstrate the existence of pairs of synchronous orbits. 

It is in fact not hard to demonstrate the existence of isofrequency orbits without resorting to a numerical calculation as above. The argument follows from a few simple observations, which we now describe. First, it is easily established that, in the $(e,\Omega_{\varphi})$ plane, the separatrix $e=e_s(\Omega_{\varphi})$ is a curve of a {\em positive} slope as shown in Fig.\ \ref{fig:Schwarzschild_Omega_phi_e} 
(noting that in the figure we have chosen the horizontal axis with $\Omega_\varphi$ increasing to the {\em left}, so that, e.g, the radius of circular orbits increases to the right) . To see this, use Eq.\ (\ref{eq:Schwarzschild_phi_frequency}) with (\ref{DeltaphiAsy}) and (\ref{TrAsy}) to derive the relation $\Omega_\varphi(e)$ along the separatrix, and invert to obtain 
\begin{eqnarray}
	e_s(\Omega_\varphi) = \frac{6\tilde\Omega_\varphi^{2/3}-1}{1-2\tilde\Omega_\varphi^{2/3}}\, ,	\label{eq:e_s}
\end{eqnarray}
where, recall, $\tilde\Omega_{\varphi}:=M\Omega_{\varphi}$.
This gives $de_s/d\Omega_{\varphi}>0$ in the relevant range $0<\tilde\Omega_{\varphi}<1/8$. Next, examine the curve $\Omega_r=0$ in the $(e,\Omega_{\varphi})$ plane: It runs up along the separatrix, then proceeds horizontally along the line $e=1$ (which represents orbits with $r_{\rm a}\to\infty$ and hence $T_r\to\infty$), and finally descends along the line $\Omega_{\varphi}=0$ (which represents weak-field orbits with $r_{\rm p}\to\infty$, for which both frequencies vanish). Hence, the $\Omega_r=0$ contour is represented by the thick red line in Fig.\ \ref{fig:Schwarzschild_Omega_phi_e}, circumscribing the parameter space of bound orbits on 3 sides. From continuity, it is now clear that a contour line of sufficiently small $\Omega_r$ must ``bend backward'' inside the wedge formed by the separatrix and the $e=1$ line, so that it becomes vertical at a point. The existence of isofrequency pairs follows immediately, as discussed above.

Let us now delineate the region in the parameter space where isofrequency pairing occurs. In Fig.\ \ref{fig:Schwarzschild_Omega_phi_e} we have indicated in a dotted black line the curve along which the transformation $(p,e)\leftrightarrow(\Omega_r,\Omega_\varphi)$ becomes singular. Each and every orbit left of this singular curve has an isofrequency dual right of the curve. In particular, each and every circular ($e=0$) orbit on the open segment $(i,b)$ has an isofrequency dual on the dashed green line marked as {\it circular-orbit duals} (COD). (Here we define the radial frequency of a circular orbit to be that of a slightly eccentric orbit, at the limit $e\to 0$.) Hence, each and every orbit between the separatrix and the singular curve has an isofrequency dual between the singular curve and the COD, and vice versa. We conclude that (i) all isofrequency pairs are confined to the region left of the COD, and (ii) every orbit left of the COD has an isofrequency dual. 

%To conclude our discussion of synchronous orbits in Schwarzschild spacetime we shall briefly mention a few of the interesting features of Fig.~\ref{fig:Schwarzschild_Omega_phi_e}. Firstly we have marked on the $J=0$ line where the transformation between $(p,e)\leftrightarrow(\Omega_r,\Omega_\varphi)$ is singular. We also mark on the curve of the circular orbit dual (COD) orbits. Orbits with parameters along this curve have a dual orbit with $e=0$. These two curves are of interest because orbits with parameters to the left-hand-side of the $J=0$ curve have a dual orbit with parameters between the $J=0$ and COD curves. We also remark that orbits with parameters to the right of the COD curve cannot be part of a synchronous orbit pair.

How ``strong field'' is the region left of the COD, where isofrequency pairing occurs? The isofrequency pair of lowest azimuthal frequency sits where the singular curve intersects the $e=0$ axis, at point $b$ (refer again to Fig.\ \ref{fig:Schwarzschild_Omega_phi_e}). To calculate the value of $\Omega_{\varphi}$ at $b$, we analytically Taylor-expand the Jacobian determinant $J:=\left|\partial(\Omega_{r},\Omega_{\varphi})/\partial(p,e)\right|$ in $e$ about $e=0$ (for fixed $p$). We find, to leading order,
\begin{equation}\label{J}
J(e\to 0)=-\frac{9(4p^2-39p+86)}{4M^2 p^{9/2}(p-2)(p-6)^{3/2}},
\end{equation}
of which the relevant root is 
\begin{equation}
p=\frac{1}{8}(39+\sqrt{145}) \simeq 6.3802. 
\end{equation}
This corresponds to a circular orbit of radius $r_b\simeq 6.3802M$ and frequency $\Omega_{\varphi}=(M/r_b^3)^{-2}\simeq 0.06205/M $. Recall this is the {\it lowest} frequency of any isofrequency pair. The isofrequency pair of {\it highest} frequency sits at the upper-left corner of the diagram in Fig.\ \ref{fig:Schwarzschild_Omega_phi_e}; it has $\tilde\Omega_\varphi=1/8$. Hence, for a Schwarzschild black hole, the range of isofrequency pairing is given by 
\begin{equation}
0.06205\lesssim \tilde\Omega_\varphi< 0.125 .
\end{equation}
(For comparison, the ISCO frequency is $\tilde\Omega_{\varphi}=6^{-3/2}\simeq 0.068$.) Evidently, the phenomenon is confined to the very strong-field regime of the Schwarzschild black hole. 

%By performing Taylor expansions of Eqs.~\ref{eq:Schwarzschild_frequencies} about $e=0$ it is straightforward to calculate where the $J=0$ curve intersects the line $e=0$. The circular orbit at the intersection has a radius of $r_b=(39+\sqrt{145})/8M$. This particular orbit is of interest as it has the smallest $\Delta\varphi$ of any orbit which is part of a synchronous orbit pairing. 

Finally, we note that all orbits in isofrequency pairs are strongly zoom-whirling. For example, the lowest-frequency isofrequency pair mentioned above (slightly perturbed circular orbits of radii $r\to r_b^{\pm}$) have $\Delta\varphi \simeq 4.1\times 2\pi$, i.e., they each complete more than 4 full revolutions in $\varphi$ over a single radial period. This behavior is also manifest in the example shown in Fig.\ \ref{fig:Schwarzschild_sample_pair}.

\section{Isofrequency orbits in Kerr geometry}\label{Sec:Kerr}

\subsection{Equatorial orbits}

We consider first the case of equatorial orbits, in which the treatment is entirely analogous to that of orbits in Schwarzschild spacetime. Equatorial orbits have $\Q=0$, and are therefore parametrized by the pair $\{\en,\ang\}$ alone. As in the Schwarzschild case, bound equatorial orbits may instead be parametrized by the (BL coordinate values of the) turning points $\{r_{\rm a},r_{\rm p}\}$, or by a pair $\{p,e\}$ defined from them as in Eq.\ (\ref{pe}). One can then write integral expressions analogous to Eqs.\ (\ref{eq:Schwarzschild_r_frequency}) [with (\ref{eq:T_r}) and (\ref{eq:dt_dchi})] and (\ref{eq:Schwarzschild_phi_frequency}) [with (\ref{eq:DeltaPhi})] for the radial and azimuthal frequencies of the motion; the dependence upon the black hole's spin $a$ only enters via the explicit form of the functions $dt/d\chi(\chi;p,e,a)$ and $d\varphi/d\chi(\chi;p,e,a)$, which are significantly more complicated than their Schwarzschild ($a=0$) reductions. The integral formulas for $\Omega_r$ and $\Omega_{\varphi}$, for arbitrary spin, can be found in Sec.\ II.A of Ref.\ \cite{Glampedakis-Kennefick}, and an analytic formula for the separatrix curve, $p_s(e)$, again for arbitrary spin, is given in Ref.\ \cite{Levin-Perez-Giz}. We will not reproduce these expressions here given their complexity, and since we will be giving explicit formulas for generic orbits in the next subsection.

%For orbits in the equatorial plane of a Kerr black hole the mapping $(p,e) \leftrightarrow (\Omega_r,\Omega_\varphi)$ is still not globally one-to-one. The equivalent equations to Eqs.~\eqref{eq:dphi_dchi} and \eqref{eq:dt_dchi} are given by Glampedakis and Kennefick \cite{Glampedakis-Kennefick} (we do not repeat them here for brevity). 
%Calculation of the separatrix curve, $p_s(e)$, for orbits in the equatorial plane can be achieved using the analytic formula given by Levin and Perez-Giz \cite{Levin-Perez-Giz}. 

One finds that our intuitive argument for the existence of isofrequency orbits carries over directly from the Schwarzschild case to equatorial orbits in Kerr. Along the separatrix of the Kerr black hole, the function $e_s(\Omega_\varphi)$ is most neatly expressed in terms of the periastron radius $\tilde r_{\rm p}=(\tilde\Omega_\varphi^{-1}-\tilde a)^{2/3}$ (which, on the separatrix, corresponds to the radius of an unstable circular orbit of frequency $\Omega_\varphi$) \cite{Levin-Perez-Giz}:
\begin{equation}
e_s =
\frac{-\tilde r_{\rm p}^2+6\tilde r_{\rm p}-8\tilde a \tilde r_{\rm p}^{1/2}+3\tilde a^2 }{\tilde r_{\rm p}^2-2\tilde r_{\rm p}+\tilde a^2}\, .    \label{eq:e_s_Kerr}
\end{equation}
It can be easily checked that $de_s/d\tilde r_w<0$ and $d\tilde r_w/d\tilde\Omega_{\varphi}<0$ for all $a$ and all $\Omega_{\varphi}$ in the relevant range $0<\Omega_{\varphi}<\Omega_{\varphi}^{\rm max}$, leading, again, to  $de_s/d\Omega_\varphi>0$.
[Here $\Omega_{\varphi}^{\rm max}$ is the whirl frequency of the marginally bound and marginally stable orbit with ${\cal E}=1$ (and $e=1$), an expression for which will be given in Eq.\ (\ref{Omegamax}) below.] The pattern of the $\Omega_r={\rm const}$ contour lines in the $(e,\Omega_\varphi)$ plane should therefore be qualitatively as in Fig.\ \ref{fig:Schwarzschild_Omega_phi_e}, including the crucial feature that contour lines  ``curve back'' inside the wedge formed by the separatrix and the $e=1$ line. It follows that isofrequency pairing should be a feature of equatorial orbits for any black hole spin $a$ (and, in particular, we expect to see it in both prograde and retrograde orbits).

\begin{figure}
	\includegraphics{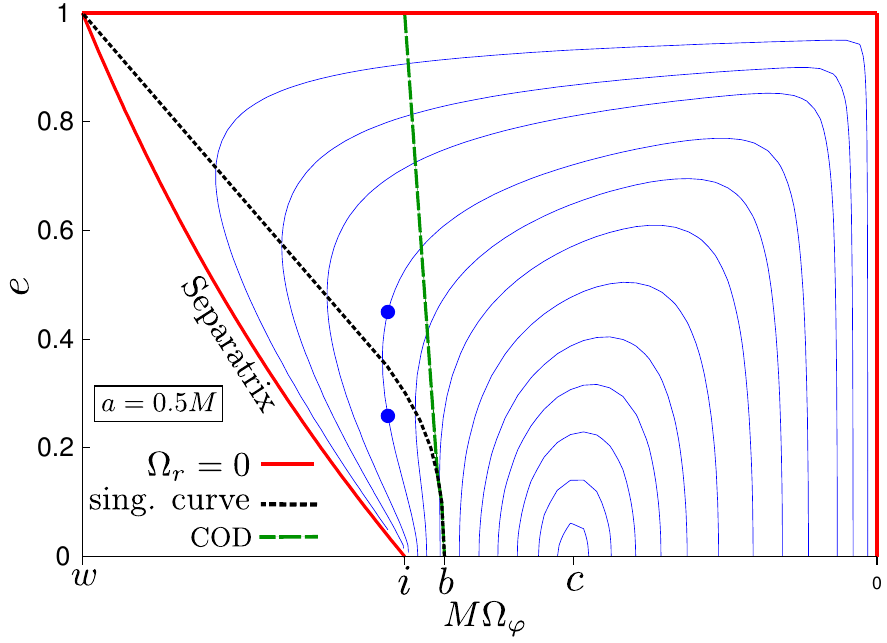}
	\caption{
	The $(\Omega_\varphi,e)$ parameter space for bound equatorial geodesic orbits in Kerr geometry with $a=0.5M$. Compare with Fig.\ \ref{fig:Schwarzschild_Omega_phi_e}. The relevant features are as in the Schwarzschild case, and the existence of isofrequency pairing below the COD is similarly evident. We indicate a sample pair with $(p,e)=(4.915656,0.45)$ and $(4.62288270,0.26313140)$, both having frequencies $(\Omega_\varphi,\Omega_r)=(0.112675037,0.01291945)$. Labelled points on the horizontal axis correspond to circular orbits of radii (left to right) $r_w\simeq 3.8994M$ (whirl radius of marginally bound marginally stable orbit; orbit of highest azimuthal frequency), $r_{i}\simeq 4.2330M$ (ISCO), $r_b\simeq 4.5039M$ (outermost orbit in an isofrequency pair), and $r_c\simeq 5.7628M$ (orbit of highest radial frequency, $M\Omega_r\simeq 0.03312$). 
	}\label{fig:KerrEq_Omega_phi_e}
\end{figure}

Figure \ref{fig:KerrEq_Omega_phi_e} shows an actual contour-line map, similar to that in Fig.\ \ref{fig:Schwarzschild_Omega_phi_e}, for the sample case $a=0.5M$. The $\Omega_r={\rm const}$ contours were computed numerically as in the Schwarzschild case, this time using the integral expressions from Ref.\ \cite{Glampedakis-Kennefick}. Near the separatrix we have used the asymptotic expressions also given in \cite{Glampedakis-Kennefick}. Evidently, the essential features are as in the Schwarzschild case. One again identifies a singular curve and a COD curve in the ($e,\Omega_\varphi$) plane, so that for any orbit between the separatrix and the singular curve there exists a dual isofrequency orbit between the singular curve and the COD, and vice versa. The situation is qualitatively the same for other values of the spin and for retrograde orbits.

%By calculating $e_s(\Omega_\varphi)$ as was done for Eq.~\eqref{eq:e_s} one finds that the separatrix maintains the same slope for all spin values of both prograde orbits (whose orbital spin is aligned with the black hole's spin) and retrograde orbits (whose orbital spin is anti-aligned), with the possible exception of the extremal ($a=M$) prograde case where the separatrix becomes a vertical line at $\Omega_\varphi=0.5$. As the separatrix maintains the same slope the argument given above for the existence of synchronous orbits carries through. Explicit  numerical calculation of $\Omega_r$ contours throughout the parameter space can be achieved using the near separatrix expansions found in Ref.~\cite{Glampedakis-Kennefick}.

\begin{figure}
	\includegraphics[width=9cm]{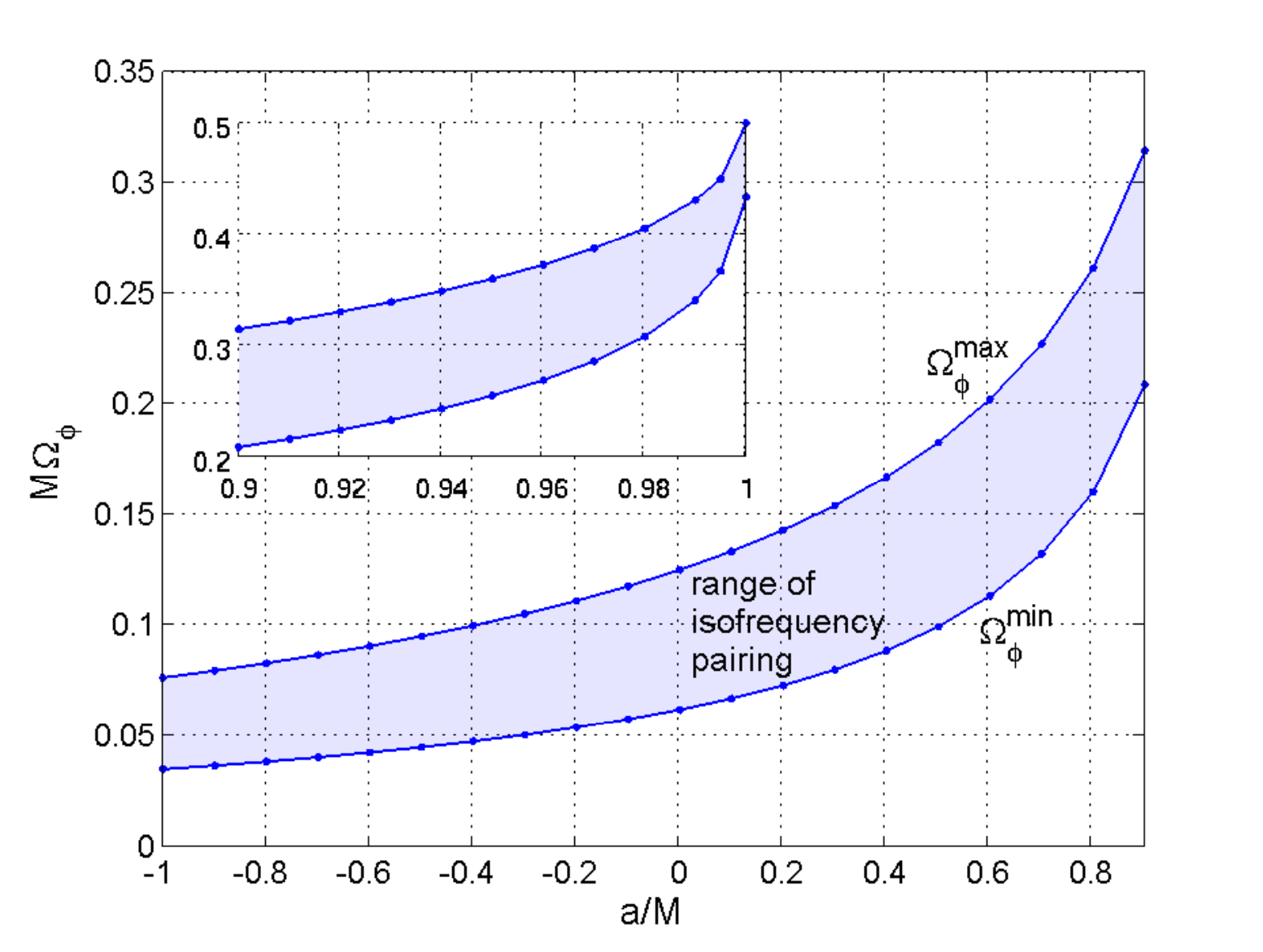}
	\caption{Range of isofrequency pairing (shaded area), as a function of the black hole spin, for orbits in the equatorial plane. The large-spin portion of the plot is shown separately in an inset for clarity. Isofrequency orbits are confined to the strong-field frequency regime $\Omega_{\varphi}>\Omega_{\varphi}^{\rm min}$. The frequency $\Omega_{\varphi}^{\rm max}$ is the highest attainable by {\it any} bound orbit (at given $M,a$), corresponding to the whirl frequency of the marginally bound, marginally stable orbit with $\en=1$ (which is also the azimuthal frequency of the ``unstable'' circular orbit with that energy).
	}\label{fig:Omegaminmax}
\end{figure}

Let us identify the frequency range $\Omega_{\varphi}^{\rm min}(a) < \Omega_{\varphi} < \Omega_{\varphi}^{\rm max}(a)$ where isofrequency pairing occurs. The $a\ne 0$ version of Eq.\ (\ref{J}) is too complicated to be solved analytically for $p=r_b$ (the radius of the outermost circular orbit belonging to an isofrequency pair) as we have done in the Schwarzschild case, so we resort to numerical solutions. Table \ref{table} lists $r_b$ values for a sample of black hole spins. Once a numerical value for $r_b$ is at hand (for a given $a$), $\Omega_{\varphi}^{\rm min}$ is obtained via 
\begin{equation}\label{Omegamin}
\tilde\Omega_{\varphi}^{\rm min}=\frac{1}{\tilde r_b^{3/2}+\tilde a},
\end{equation}
where we have used the general relation between the frequency of a circular equatorial orbit and its BL radius \cite{Bardeen-Press-Teukolsky}.
%Table \ref{table} lists $r_b$ values for a sample of black hole spins, and in Fig.\ \ref{fig:Omegaminmax} we plot $\Omega_{\varphi}^{\rm min}$ as a function of $a$. 
The {\it maximal} value $\Omega_{\varphi}^{\rm max}$ corresponds to the whirl frequency of the marginally bound marginally stable orbit with $e=1$ (top left corner in Fig.\ \ref{fig:KerrEq_Omega_phi_e}). It is given by 
\begin{equation}\label{Omegamax}
\tilde\Omega_{\varphi}^{\rm max}=
\frac{1}{(2-\tilde a+2\sqrt{1-\tilde a})^{3/2}+\tilde a}.
\end{equation}
%where $\tilde r_w=2-\tilde a+2\sqrt{1-\tilde a}$ is the whirl radius.
The range $\Omega_{\varphi}^{\rm min}(a) < \Omega_{\varphi} < \Omega_{\varphi}^{\rm max}(a)$ is illustared in Fig.\ \ref{fig:Omegaminmax}.

%----------------------------------------------------------------------------------------
\begin{table}[htb] 
\begin{tabular}{|c|cc|}
\hline\hline
$\tilde a$ & $\tilde r_{\rm isco}$ & $\tilde r_{b}$ \\
\hline\hline
0	&	6		&	6.38020		\\
0.1 &	5.66930	&	6.02903	    \\	
0.2 &	5.32944	&	5.66813	    \\
0.3 &	4.97862	&	5.29559	    \\
0.4 &	4.61434	&	4.90877	    \\
0.5 &	4.23300	&	4.50387	    \\
0.6 &	3.82907	&	4.07499	    \\
0.7 &	3.39313	&	3.61219	    \\
0.8 &	2.90664	&	3.09586	    \\
0.9 &	2.32088	&	2.47458	    \\
0.95 &	1.93724	&	2.06835	    \\
0.99 &	1.45450	&	1.56060	    \\
1 	&	1		&	1.19441	    \\
\hline\hline
\end{tabular}
\quad\quad 
\begin{tabular}{|c|cc|}
\hline\hline
$\tilde a$ & $\tilde r_{\rm isco}$ & $\tilde r_{b}$ \\
\hline\hline
-0.1 &	6.32289	&	6.72309	    \\	
-0.2 &	6.63904	&	7.05292	    \\
-0.3 &	6.94927	&	7.38801	    \\
-0.4 &	7.25427	&	7.71208	    \\
-0.5 &	7.55458	&	8.03103	    \\
-0.6 &	7.85069	&	8.34549	    \\
-0.7 &	8.14297	&	8.65588	    \\
-0.8 &	8.43176	&	8.96255	    \\
-0.9 &	8.71735	&	9.26583	    \\
%
%-0.99 &	8.97186	&	9.53610       \\
%
-1 	&	9		&  	9.56598       \\
\hline\hline
\end{tabular}

\caption{
Numerical values for $r_b$, the BL radius of the outermost circular orbit belonging to
an isofrequency pair (cf.\ Fig.\ \ref{fig:KerrEq_Omega_phi_e}). The frequency $\Omega_{\varphi}^{\rm min}$ of this orbit [given in Eq.\ (\ref{Omegamin})] marks the lower end of the frequency range where synchronous pairing occurs. For comparison, the second column displays the ISCO radius $r_{\rm isco}$ (elsewhere in this paper denoted $r_i$); it is given by \citep{Bardeen-Press-Teukolsky} 
$\tilde r_{\rm isco}=3+Z_2-{\rm sign}(a)[(3-Z_1)(3+Z_1+2Z_2)]^{1/2}$, where 
$Z_1:=1+(1-\tilde a^2)^{1/3}[(1+\tilde a)^{1/3}+(1-\tilde a)^{1/3}]$ and $Z_2:=(3\tilde a^2+Z_1^2)^{1/2}$.
Numerical values are truncated at the 5th decimal place, rounding up.
}
\label{table}
\end{table}
%----------------------------------------------------------------------------------------

\subsection{Triperiodic orbits: frequencies and separatrix}\label{sec:Kerr_freqs}

We now turn to consider generic bound motion in Kerr geometry. Nonequatorial orbits possess a third frequency, $\Omega_\theta$, associated with the longitudinal motion. It is not immediately obvious how the three fundamental frequencies can be computed in practice, since the radial and longitudinal motions are coupled in the usual BL-coordinate representation [see Eqs.\ (\ref{EOMr}) and (\ref{EOMq}) below]. Schmidt \cite{Schmidt} was able to derive formal expressions for the fundamental frequencies using angle-action variables in the Hamilton--Jacobi formalism, which circumvented the problem of coupling. Mino \cite{Mino} observed that the radial and longitudinal motions can in fact be decoupled using a simple transformation of the time coordinate, and Fujita and Hikida \cite{Fujita-Hikida} (building on work by Drasco and Hughes \cite{Drasco-Hughes}) used this to obtain closed-form analytic formulas for the three frequencies. We give their formulas below in a slightly modified form. (Fujita and Hikida considered the cases $|a|\ne M$ and $|a|=M$ separately. For brevity we reproduce here only the nonextremal case; expressions for $|a|=M$ can be found in Appendix B of \cite{Fujita-Hikida}.)

To establish some necessary notation, let us begin with the $r$ and $\theta$ components of the geodesic equation of motion. For bound (${\cal E}<1$), nonequatorial ($\theta_{\min}\ne \pi/2$) orbits around a rotating ($a\ne 0$) black hole, these can be written in the form  
\begin{equation}\label{EOMr}
\Sigma^2 \dot r^2 = \gamma(r_1-r)(r-r_2)(r-r_3)(r-r_4), 
\end{equation}
\begin{equation}\label{EOMq}
\Sigma^2 \dot z^2 = a^2 \gamma(z_-^2-z^2)(z_+^2-z^2),
\end{equation}
where $z:=\cos\theta$, $\Sigma:=r^2+a^2 z^2$, $\gamma:=1-{\cal E}^2$, and an overdot denotes differentiation with respect to proper time along the geodesic. The roots of the quartic expressions on the right-hand sides are certain functions of ${\en,\ang,\Q}$; the radial roots are ordered as $r_1\geq r_2\geq r_3\geq r_4$, and the roots $\pm z_-,\pm z_+$ satisfy $|z_-|\leq 1$ and $|z_+|> 1$.  Bound orbits have $r_{\rm p}\equiv r_2\leq r\leq r_1\equiv r_{\rm a}$ and $|z|\leq z_-\equiv \cos\theta_{\min}$ (the latter inequality corresponds to $\theta_{\min}\leq\theta\leq \pi-\theta_{\min}$). We may introduce the parametrization $\{p,e,\theta_{\min}\}$, where $p,e$ are defined from $r_{\rm p},r_{\rm a}$ as in Eq.\ (\ref{pe}). The above roots are then most succinctly expressed (using a ``mixed'' parametrization) as
\begin{equation}\label{zpm}
z_- = \cos\theta_{\min},\quad\quad
z_+=\left(1+\frac{\ang^2}{a^2\gamma\sin^2\theta_{\min}}\right)^{1/2},
%z_+^2=\frac{{\cal Q}}{a^2(1-{\cal E}^2)z_-^2},
\end{equation}
\begin{equation}\label{r12}
r_1\equiv r_{\rm a}= \frac{Mp}{1-e}, \quad\quad  r_2\equiv r_{\rm p}= \frac{Mp}{1+e},
\end{equation}
\begin{equation}\label{r34}
	r_3 = \frac{1}{2}\left[\alpha + \sqrt{\alpha^2 - 4\beta}\right]\,,\qquad r_4 = \frac{\beta}{r_3},
\end{equation}
where
$\alpha 	:= 2M/\gamma - (r_{\rm a}+r_{\rm p})$ and
$\beta   := a^2\mathcal{Q}/(\gamma r_{\rm a}r_{\rm p})$.

Note that the $r$ and $\theta$ motions are coupled, due to the factor $\Sigma^2(r,\theta)$ on the left-hand sides of Eqs.\ (\ref{EOMr}) and (\ref{EOMq}). This can be easily rectified by introducing a new time parameter $\lambda$ (often referred to as ``Mino time'' in recent literature), satisfying $\dot\lambda=\Sigma^{-1}$. In terms of $\lambda$, the $r$ and $\theta$ motions decouple, and each becomes manifestly periodic, with $\lambda$-frequencies $\Upsilon_r$ and $\Upsilon_\theta$, respectively. One can also define the azimuthal frequency $\Upsilon_\varphi$ as the average of $d\varphi/d\lambda$ with respect to $\lambda$, where in general the average needs to be taken over an infinite time. The three $\lambda$-frequencies are given explicitly (for $|a|\ne M$) by \cite{Fujita-Hikida}
\begin{eqnarray}
	\Upsilon_r 		&=& \frac{\pi\sqrt{\gamma(r_{\rm a}-r_3)(r_{\rm p}-r_4)}}{2K(k_r)}\,,		\label{eq:Upsilon_r}	\\
	\Upsilon_\theta &=& \frac{\pi (a^2 \gamma)^{1/2} z_+}{2K(k_\theta)}\,,									\\
	\Upsilon_\varphi	&=&	\frac{\ang\Pi(z_-^2,k_\theta)}{K(k_\theta)} + \frac{a}{r_+ - r_-}\left[\frac{2M\en r_+ - a\ang}{r_3 - r_+} \right. \label{eq:Upsilon_phi} \\
					&\times& \left. \left( 1 - \frac{\mathcal{F}_+}{r_{\rm p}-r_+}\right) - (+ \leftrightarrow -) \right]\,, \nonumber
\end{eqnarray}
where $\Pi(x,y):=\int_{0}^{\pi/2}d\theta(1-x\sin^2\theta)^{-1}(1-y\sin^2\theta)^{-1/2}$ is the complete elliptic integral of the third kind, $r_\pm:=M\pm\sqrt{M^2-a^2}$, the arguments of the elliptic functions are
\begin{eqnarray}
	k_r := {\frac{r_{\rm a}-r_{\rm p}}{r_{\rm a}-r_3}\frac{r_3-r_4}{r_{\rm p}-r_4}}\,,\qquad k_\theta := (z_-/z_+)^2,
\end{eqnarray}
and hereafter we use $(+ \leftrightarrow -)$ to denote a term formed by interchanging the $+$ and $-$ subscripts in the previous terms within the enclosing brackets. In Eq.\ (\ref{eq:Upsilon_phi}) we have also introduced 
\begin{equation}
	\mathcal{F}_A := (r_{\rm p} - r_3) \frac{\Pi(h_A,k_r)}{K(k_r)} 
\end{equation}
for $A=\{r,+,-\}$, with 
\begin{eqnarray} \label{hpm}
	h_\pm = \frac{(r_{\rm a}- r_{\rm p})(r_3 - r_\pm)}{(r_{\rm a} - r_3)(r_{\rm p} - r_\pm)}\,,\qquad h_r = \frac{r_{\rm a} - r_{\rm p}}{r_{\rm a}-r_3}\,.
\end{eqnarray}

Finally, the $t$-frequencies are obtained from the $\lambda$-frequencies via \cite{Drasco-Hughes}
\begin{eqnarray}
	\Omega_r = \frac{\Upsilon_r}{\Gamma}\,,\qquad \Omega_\theta = \frac{\Upsilon_\theta}{\Gamma}\,,\qquad \Omega_\varphi = \frac{\Upsilon_\varphi}{\Gamma}\,,\label{eq:freqs_from_mino_to_t}
\end{eqnarray}
where $\Gamma$ is the average of $dt/d\lambda$ with respect to $\lambda$. The latter is given explicitly (for $|a|\ne M$) by \cite{Fujita-Hikida}
\begin{widetext}
\begin{eqnarray} \label{Gamma}
	\Gamma &=& 4M^2\mathcal{E} + \frac{\mathcal{E}\mathcal{Q}(1-\mathcal{G}_\theta)}{\gamma z_-^2} + \frac{\mathcal{E}}{2}\left[r_3(r_{\rm a}+r_{\rm p}+r_3) - r_{\rm a} r_{\rm p} + (r_{\rm a} + r_{\rm p} + r_3 + r_4)\mathcal{F}_r +(r_{\rm a}-r_3)(r_{\rm p}-r_4)\mathcal{G}_r\right]	\hspace{0.5cm}	\\
			&+& 2M\mathcal{E}(r_3+\mathcal{F}_r) + \frac{2M}{r_+ - r_-}\left[ \frac{(4M^2\mathcal{E} - a\mathcal{L})r_+ - 2Ma^2\mathcal{E}}{r_3 - r_+}\left(1-\frac{\mathcal{F}_+}{r_{\rm p}-r_+}\right) - (+\leftrightarrow -) \right]\,, 		\nonumber
\end{eqnarray}
\end{widetext}
where we have also introduced 
\begin{equation}\label{G}
 \mathcal{G}_B := \frac{E(k_B)}{K(k_B)}
\end{equation}
for $B=\{r,\theta\}$, with $E(x):=\int_{0}^{\pi/2}d\theta(1-x\sin^2\theta)^{1/2}$ being the complete elliptic integral of the second kind.

Equation (\ref{eq:freqs_from_mino_to_t}), with the necessary substitutions from Eqs.\ (\ref{zpm})--(\ref{hpm}), (\ref{Gamma}) and (\ref{G}), gives closed-form expressions for the fundamental frequencies $\Omega_r$, $\Omega_{\theta}$ and $\Omega_\varphi$, given the parameters $\{\en,\ang,\Q\}$ as well as the corresponding parameters $\{p,e,\theta_{\min}\}$. To complete the formulation, one requires a link between the two sets of parameters. Explicit expressions for $\{{\en,\ang,\Q}\}$ in terms of $\{p,e,\theta_{\min}\}$ were derived by Schmidt in Appendix B of Ref.\ \cite{Schmidt} (they are reproduced in a somewhat more concise form in Appendix A of \cite{Drasco-Hughes-2006}). With this link, Eq.\ (\ref{eq:freqs_from_mino_to_t}) can be used to compute the fundamental frequencies for a geodesic with given $\{p,e,\theta_{\min}\}$.

The {\it separatrix} between stable and unstable orbits is given by the condition $r_{\rm p}=r_3$, which identifies the point where the inner turning point of the bound orbit is lost [recall Eq.\ (\ref{EOMr})]. It can be checked that this condition coincides with $\Omega_r=0$, as expected (note $k_r=1=h_A$ and $\mathcal{F}_A =0= \mathcal{G}_r$ along the separatrix).
Using Eqs.\ (\ref{r12}) and (\ref{r34}), with the link between $\{{\cal E},{\cal L},{\cal Q}\}$ and $\{p,e,\theta_{\min}\}$ from \cite{Schmidt,Drasco-Hughes-2006}, the condition $r_{\rm p}=r_3$ translates to a relation between $p$, $e$ and $\theta_{\min}$, which can be solved numerically for $p$ to obtain the separatrix surface $p=p_s(e,\theta_{\min})$. We checked, using numerical examples, that this procedure for identifying the separatrix is consistent with the analytical method of Ref.~\cite{Levin-Perez-Giz} for equatorial orbits, and with the alternative numerical method of Sundararajan \cite{Sundararajan} for generic orbits.

%As it will be useful shortly we note that for generic orbits the separatrix can be located in the following way. First recall that along the curve $p=p_s$ we have $\Omega_r=0$. We are interested in the last stable orbit for generic, non-circular orbits and so $r_1$ is strictly greater than $r_2$ and hence $r_1-r_3\neq0$ and similarly $r_2-r_4\neq0$ (recall $r_1\ge r_2 \ge r_3 \ge r_4$). By examining Eqs.~\eqref{eq:freqs_from_mino_to_t} and \eqref{eq:Upsilon_r}, we then see $\Omega_r=0$ can only occur if $K(k_r)$ diverges i.e., when $k_r=1$ . This in turn implies that at the separatrix we have $r_3-r_2=0$. Solving this equation for fixed $e$ and $\theta_\text{min}$ gives the value of $p_s(e,\theta_\text{min})$ at the separatrix. We find that this simple technique is in agreement with analytical results from Ref.~\cite{Levin-Perez-Giz} for equatorial orbits and in agreement with the numerical results of Sundararajan \cite{Sundararajan} for generic orbits. Note that at the separatrix we have $\mathcal{F}_\alpha = \mathcal{G}_r = 0$ [For $\mathcal{G}_r$ this is clear as the $K(k_r)$ term in the denominator diverges whilst the numerator remains finite. For $\mathcal{F}_\alpha$ a little more care is required but the result is straightforward to show].

\subsection{Isofrequency pairing in triperiodic orbits}

We now seek to demonstrate the existence of isofrequency pairs of triperiodic orbits, i.e., ones sharing all three fundamental frequencies $\{\Omega_r,\Omega_\theta,\Omega_\varphi\}$. Here our analysis will not be as complete as it was for biperiodic orbits. Rather, we will content ourselves with demonstrating by way of numerical example that such pairing does indeed occur.  

To this end it will suffice to inspect the contour map of $\Omega_r$=const curves in the $(e,\Omega_{\theta}$) plane, for some fixed value of $\Omega_{\varphi}$. For this, we need to be able to compute $\Omega_r$ given $\{\Omega_\theta,e,\Omega_\varphi\}$. To achieve this in practice we take the following steps. First, we numerically invert, for given $e,\theta_{\min}$, the equation $\Omega_\varphi(p)=\text{const}$ (in the example presented below we take the constant to be $0.14M^{-1}$). For this we use a bisection method, taking as initial guess the value  $p=p_s(e,\theta_{\min})$ obtained using the method described above. Once we have the trio $\{p,e,\theta_{\min}\}$, we calculate the corresponding values of $\Omega_\theta$ and $\Omega_r$ using the analytic expressions presented above. We repeat these two steps for a great many values of $e$ and $\theta_{\min}$, making sure to achieve a good coverage of the parameter space, particularly near the separatrix. %To ensure that we get a good coverage of the parameter space, particularly near the separatrix, we have used the following procedure. For a given $e$ we increase $\theta_{\min}$ by a (fixed) small increment. After each increase we check to see if the new values of $p,e,\theta_{\min}$ correspond to a bound orbit. If not we return to the previous value of $\theta_{\min}$ and increment it by a smaller amount, repeating this process until we are sufficiently close to the separatrix. 
The outcome of this procedure is a list of $\{\Omega_\theta,e,\Omega_r\}$ values for many orbits, all with our fixed value of $\Omega_\varphi$. This dataset can then be used to create a contour map of $\Omega_r$=const curves in the $(e,\Omega_{\theta}$) plane.

We remark that the analytic formulation by Fujita and Hikida proves extremely useful for our purpose, because is can be readily implemented on a computer algebra platform such as \textit{Mathematica}, which allows for high precision floating-point arithmetic. In our procedure, such high precision is crucial near the separatrix, and it avoids the need to use asymptotic expansions as in Eqs.\ (\ref{DeltaphiAsy}) and (\ref{TrAsy}).

%formulae in the above algorithm greatly reduces the complexity of the calculation (over, say, using Schmidt's formulae) as no expansions in $\epsilon$ near the separatrix are required; Fujita and Hikida's formulae are easy to work with throughout the parameter space of bound orbits using computational packages such as \textit{Mathematica} which are extremely efficient at computing elliptic integrals.

\begin{figure}
	\includegraphics[width=85mm]{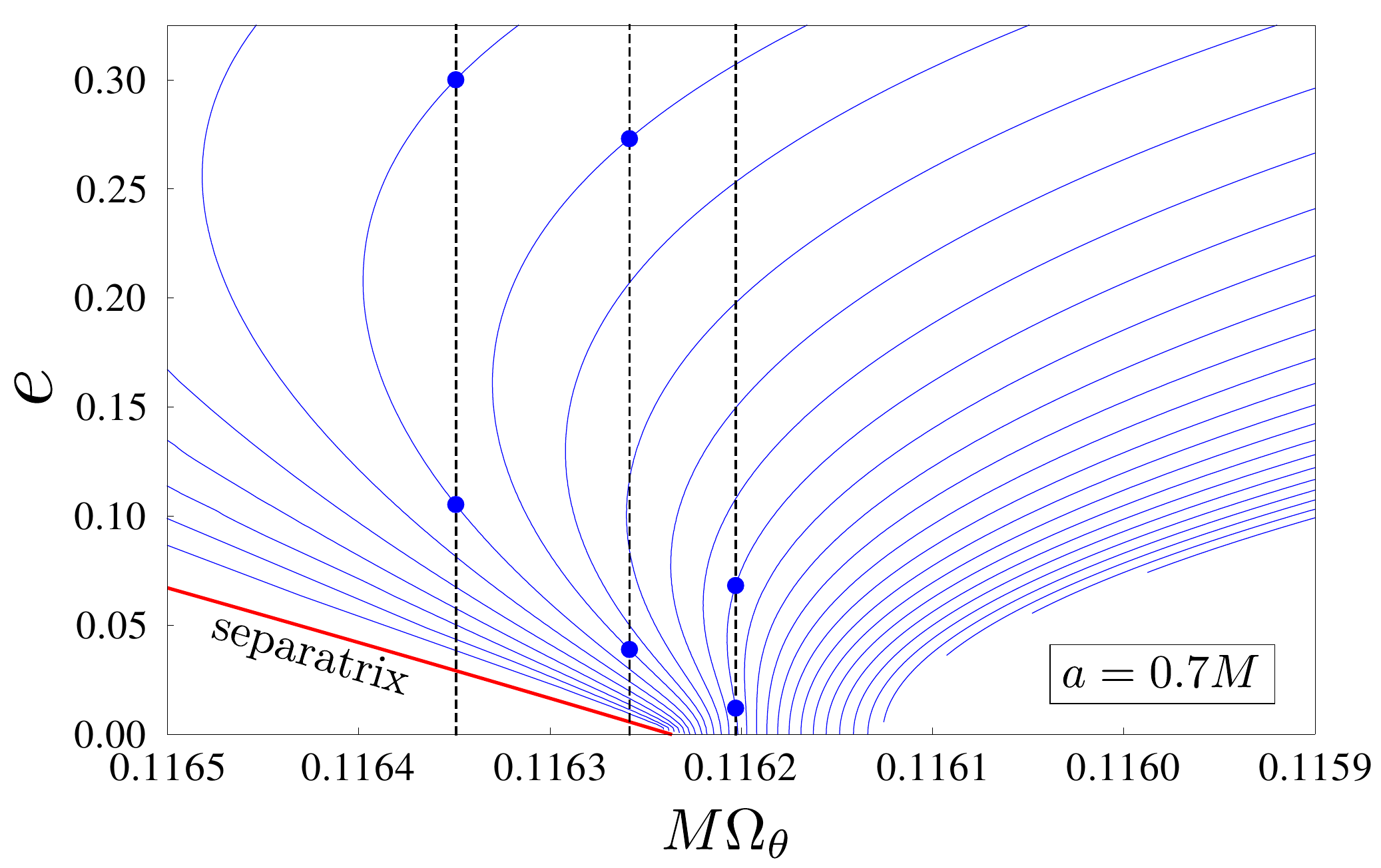}
	\caption{Illustration of isofrequency pairing in triperiodic orbits. Thin solid (blue) lines are contours of constant $\Omega_r$ in the $(e,\Omega_{\theta})$ plane, for inclined eccentric orbits with fixed $\tilde\Omega_\varphi = 0.14$. Here $a=0.7M$. (The ``empty'' lower-right corner of the diagram lies outside the parameter space of bound orbits.) Dashed vertical lines are sample $\Omega_\theta$=const contours, along each of which we indicate a pair of isofrequency orbits. The parameters of these three pairs are given in Table \ref{table:kerr_sample_pairs} (sample pairs `1', `2' and `3', from right to left). All essential features are as in Figs.~\ref{fig:Schwarzschild_Omega_phi_e} and \ref{fig:KerrEq_Omega_phi_e}. Similar contour maps can be obtained for other values of $\tilde\Omega_\varphi$ and $a$.
	}\label{fig:Kerr_case}
%As with the Schwarzschild case (cf.\ Fig.~\ref{fig:Schwarzschild_Omega_phi_e}) the shape of the $\Omega_r$ contours implies that the $\Omega_\theta$ contours intersect them twice in certain regions of the parameter space. As all points shown also have a fixed $\Omega_\varphi$ this demonstrates the existence of pairs of orbits that share the same three orbital frequencies. 
%Lastly we note that very close to the separatrix it is difficult to numerically evaluate the orbital frequencies and for this reason we show no data in this region. The region at the bottom left is also left empty as there are no orbits with these parameters.}
\end{figure}

An example with $\tilde\Omega_\varphi=0.14$ is shown in Fig.~\ref{fig:Kerr_case}. We observe that the essential features of the contour map are just as in Figs.\ \ref{fig:Schwarzschild_Omega_phi_e} and \ref{fig:KerrEq_Omega_phi_e}. In particular, there are vertical ($\Omega_{\theta}$=const) lines that cross single $\Omega_r$=const contours twice. Each pair of intersections represents a pair of isofrequency orbits sharing all three frequencies $\{\Omega_r,\Omega_\theta,\Omega_\varphi\}$. The existence of isofrequency pairing in triperiodic orbits is thus established. Continuity suggests that there should be a certain volume in the 3-dimensional parameter space where isofrequency orbits reside, but here we will not endeavour to identify the boundaries of this volume.

\begin{figure}
	\includegraphics[width=85mm]{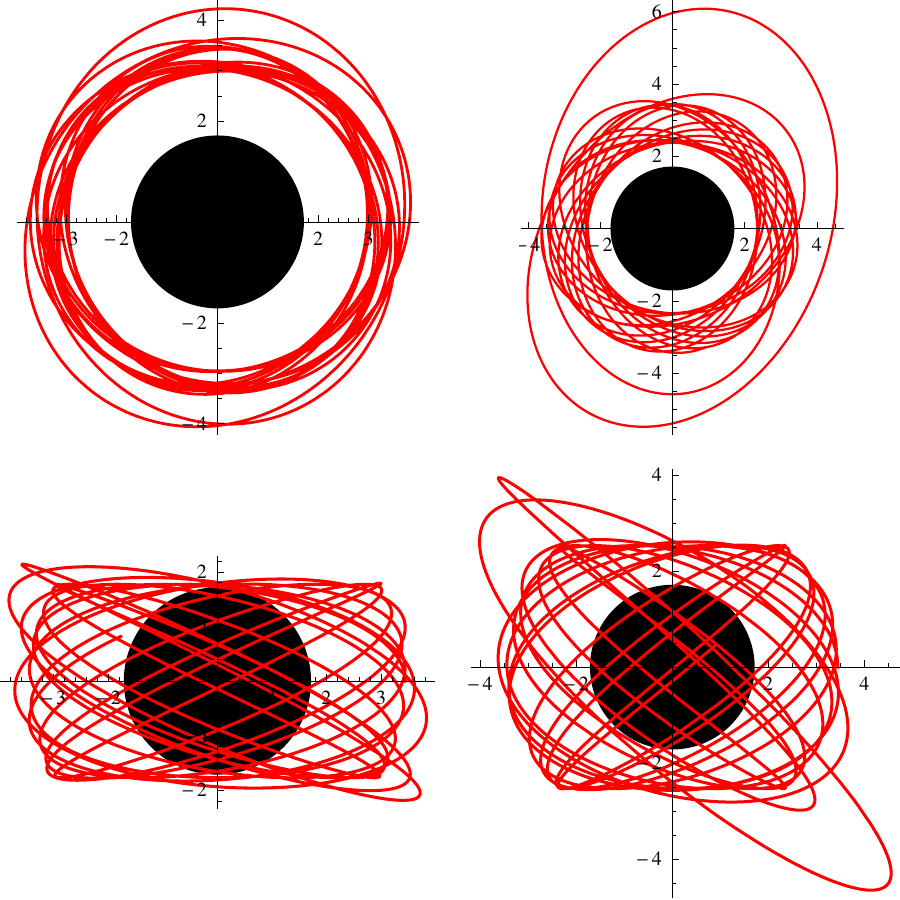}
	\caption{A sample pair of triperiodic isofrequency orbits for $a=0.7M$. The orbits depicted correspond to `sample pair 3' from Table \ref{table:kerr_sample_pairs} (also leftmost pair in Fig.\ \ref{fig:Kerr_case}), with `orbit 1' shown on the left and `orbit 2' shown on the right. The top row shows the motion in the $(x,y)$-plane and the bottom row shows the motion in the $(x,z)$-plane, where $x=r\cos\varphi\sin\theta/M$, $y=r\sin\varphi\sin\theta/M$ and $z=r\cos\theta/M$. The black hole is shown to scale. In both orbits the motion begins at $t=0=\lambda$ at periastron, with $\varphi=0$ and $\theta=\pi/2$. In integrating the geodesic equations we used the method of Drasco and Hughes \cite{Drasco-Hughes}, which avoids numerical difficulties near the orbital turning points. We show the portion of the orbits between $\lambda=0$ and $\lambda=30M^{-1}$. }\label{fig:Kerr_sample_pair}
\end{figure}

\begin{figure}
	\includegraphics[width=85mm]{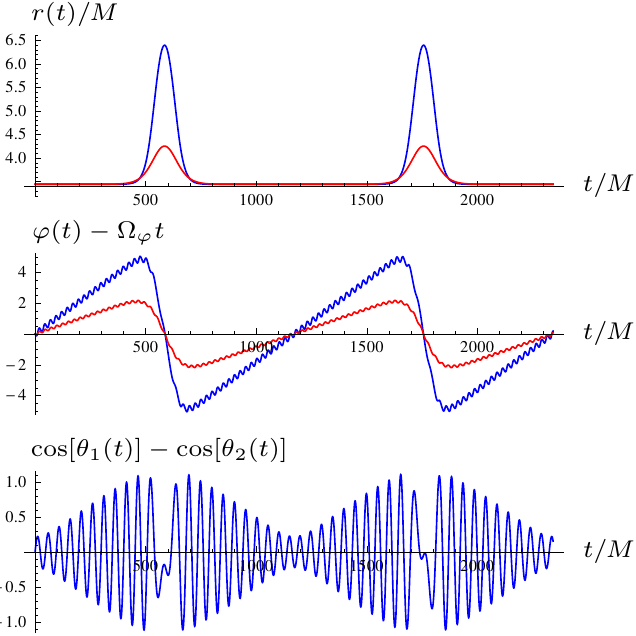}
	\caption{Evolution of $r(t)$, $\varphi(t)-\Omega_\varphi t$ and $\cos[\theta_1(t)]-\cos[\theta_2(t)]$ for `sample pair 3' of Table \ref{table:kerr_sample_pairs} and Fig.~\ref{fig:Kerr_sample_pair}. Both orbits begin at $t=0$ at periastron with $\varphi=0$ and $\theta=\pi/2$. Triperiodic isofrequency orbits are ``synchronized'' only in a long-time average sense. Periastra are reached only approximately at the same time (as a closer inspection of the upper panel would reveal) but the time differences should average to zero over a long time. The same applies to the average azimuthal motion (middle panel, where a close inspection reveals that the azimuthal phases of the two orbits are not in precise agreement at the periastra), and to the motion in $\theta$ (lower panel). In the latter case we show the {\it difference} between the two longitudinal phases, which remains quasi-periodic. It would have not remained quasi-periodic had the two orbits not been in an isofrequency pair.}\label{fig:Kerr_sample_pair_r_theta_phi}
\end{figure}

We have indicated in Fig.~\ref{fig:Kerr_case} three sample pairs of triperiodic isofrequency orbits, whose parameters we give in Table \ref{table:kerr_sample_pairs}.
%For concreteness, we give in Table \ref{table:kerr_sample_pairs} the orbital parameters of three pairs of triperiodic synchronous orbit (these are also marked in Fig.~\ref{fig:Kerr_case}). 
The third sample pair is visualized in real space in Fig.~\ref{fig:Kerr_sample_pair}, and in Fig.~\ref{fig:Kerr_sample_pair_r_theta_phi} we illustrate the synchronized evolutions of $r(t)$, $\theta(t)$ and $\varphi(t)$ for this pair. We note that in the case of triperiodic orbits the ``synchronization'' is not exact (because the $r$ and $\theta$ motions are not separately periodic). Rather, the isofrequency orbits are synchronized only in a long-time average sense. For example, if the two orbits pass their respective periastra at $t=0$, they may pass subsequent periastra at slightly different times, but the discrepancy should average to zero over a long time. One way to identify such behavior is by inspecting the difference between the orbital phases of the two orbits: the difference will remain quasi-periodic only if the two are isofrequency. The lower panel of Fig.~\ref{fig:Kerr_sample_pair_r_theta_phi} exemplifies this for the $\theta$ phase.  

\begin{table*}[htb]
    \begin{center}
        \begin{tabular}{c | c |c c ||l}
	 	&	& orbit 1 		& orbit 2 		& fundamental frequencies 					\\
	\hline\hline
	\multirow{3}{*}{sample pair 1} & $p$		& $3.615857065600587178089905$	& $3.4855158540000000000000$	& $\tilde\Omega_r=0.009040307723329$			\\
	 & $e$		& $0.068206017767752935160626$	& $0.01201000000000000000000$	& $\tilde\Omega_\theta=0.1162029753375$		\\
	 & $\theta_{\min}$	& $1.953894865146840010777339$	& $1.8378366992075975844562$	& $\tilde\Omega_\varphi=0.1400000000000$		\\
	\hline
	\multirow{3}{*}{sample pair 2} & $p$		& $3.572207717388546694585166$	& $4.3523435765502772064368261758$	& $\tilde\Omega_r=0.006051252001160$	 	\\
	 & $e$		& $0.0388932801825514054684027$	& $0.27300090000000000000000000000$	& $\tilde\Omega_\theta=0.1162584817374$	 		\\
	 & $\theta_{\min}$	& $1.943740959072074359824863$	& $2.3444136677413832042020276247$	& $\tilde\Omega_\varphi=0.1400000000000$		\\
	\hline
	\multirow{3}{*}{sample pair 3} & $p$	& $3.80671950837597698109947211$	& $4.477551959004760003175297459526$			&  $\tilde\Omega_r=0.005364669707792$			\\
	 & $e$		& $0.105336584613486946768869507$			& $0.300000000000000000000000000000$	& $\tilde\Omega_\theta=0.1163492371285$					\\
	 & $\theta_{\min}$	& $2.11092907046831122994268532$	& $2.395463898362217344327765579750$	& $\tilde\Omega_\varphi=0.1400000000000$   	\\
        \end{tabular}
        \caption{Sample pairs of triperiodic isofrequency orbits for $a=0.7M$ (cf.\ Fig.~\ref{fig:Kerr_case}). The high precision of the parameter values presented is necessary for the orbital frequencies to match to within the 13 significant figures displayed. This level of precision is required because, near the separatrix, small changes in the orbital parameters can result in comparatively large changes in the frequencies.}\label{table:kerr_sample_pairs}
    \end{center}
\end{table*}

Before concluding, let us comment on the validity of our numerical algorithm, which, as already mentioned, involves delicate high precision computation of the orbital frequencies. To establish confidence in our results we tested our code in a number of ways. First, we checked that our code reproduces all the double-precision-accurate results for $\{\en, \ang,\Q\}$ (given $\{p,e,\theta_{\min}\}$) tabulated in Ref.~\cite{Drasco-Hughes-2006}. We also verified, to over one hundred significant figures, that the results of Fujita and Hikida's orbital frequency formulas (in the form given above) agree with the results of Schmidt's less explicit formulas \cite{Schmidt}. We further validated our equations using a direct numerical integration of the $\lambda$-time geodesic equations in a few test cases. We were able to reproduce the analytically calculated $\lambda$-frequencies $\Upsilon_r$ and $\Upsilon_\theta$ to within 25 significant figures. (The quantities $\Upsilon_\varphi$ and $\Gamma$ involve infinite time averages and are therefore less easily tested in this manner.)

\section{Concluding remarks}

In this article we have shown that the three fundamental frequencies of bound geodesics in Kerr geometry do not constitute a good parametrization of the orbits in the strong-field regime. We identified a mapping between pairs of physically distinct orbits that possess the same set of orbital frequencies. A pair of isofrequency orbits are ``synchronous'' in that they exhibit the same periastron and Lense-Thirring precession rates. All orbits in isofrequency pairs are confined to the very strong-field regime near the innermost stable orbit---cf.\ Table \ref{table} and Fig.\ \ref{fig:Omegaminmax}. (Some orbits in isofrequency pairs have very large eccentricities and apastra at arbitrarily large radii, but their periastra are in the very strong field.) Our numerical experiments suggest that all members of isofrequency pairs are of ``zoom-whirl'' type, but this is yet to be checked more thoroughly in the case of triperiodic orbits and across all spin values. 

The first practical lesson from our analysis is a cautionary note for colleagues studying the data-analysis problem for gravitational-wave detectors, in particular the problem of parameter extraction for systems of extreme-mass-ratio inspirals (EMRIs). The fundamental frequencies extracted from a ``snapshot'' of an EMRI waveform, on their own, as a matter of principle, do not necessarily provide enough information from which to extract the system's intrinsic physical parameters $\en,\ang,\Q$ (or $p,e,\theta_{\min}$). If the system is sufficiently close to the innermost stable orbit, a measurement of the instantaneous frequencies could at most narrow down on two possible sets of system parameters. This ``degeneracy'', however, can be removed in any one of the following ways: (i) by examining the power spectrum of the waveform (the power distribution among the various harmonics of the fundamental frequencies will be different for the two orbits); (ii) by inspecting the waveform snippet in the time domain (the shape of the waveform is strongly dependent upon the eccentricity, for instance); or (iii) by accounting for radiation-reaction evolution effects (two orbits which are instantaneously isofrequency will evolve radiatively in different ways).   

At a more fundamental level, our analysis identifies a new feature in the strong-field dynamics of compact-object binaries in general relativity. The fundamental frequencies in a bound binary (of any mass ratio) are important invariant characteristics of the ``conservative'' sector of the dynamics. As such they have long been studied in the context of post-Newtonian (PN) theory. The instantaneous frequencies in a binary of inspiralling black holes can even, nowadays, be extracted from high-precision fully nonlinear simulations in numerical relativity (NR)---see, for example, Ref.\  \cite{LeTiec_etal-2011}. Our analysis here revealed the occurrence of isofrequency pairing in the test-particle limit (i.e., the limit of vanishing mass ratio), but it is not unreasonable to speculate that the phenomenon is a general feature of the dynamics in strongly gravitating binaries, and would reveal itself also when the mass ratio is finite. It is not clear if available PN theory can predict isofrequency pairing---this would be interesting to check. When new, higher-order PN terms are calculated in the future, it would again be interesting to check if they reveal the phenomenon, as a way of assessing the faithfulness of the PN expressions in the strong-field regime. It would also be interesting to examine whether the phenomenon manifests itself in NR simulations of inspiralling black holes of comparable masses near the innermost stable orbit. 

Because the fundamental frequencies are {\em invariant} characteristics of the conservative dynamics, they are useful as reference points for comparing the predictions of different approaches to the relativistic two-body problem. Recent examples of such ``cross-cultural'' comparisons include (i) calculations of the ISCO frequency in the self-force (SF), PN and effective-one-body (EOB) approaches \cite{Barack-Sago-PRL,Damour-2010,Favata-2011}; and (ii) calculations of the periastron advance in slightly eccentric orbits in SF, PN, EOB and NR \cite{Damour-2010,Barack-Damour-Sago,LeTiec_etal-2011}. In both examples (which involve two nonrotating black holes) relations between the two invariant frequencies associated with infinitesimally perturbed circular orbits were utilized as benchmarks for comparison. The singular curve/surface identified in our current work is an {\it invariant structure} in the parameter space \footnote{By ``invariant structure'' we refer to the fact that the singular surface in the frequency space is invariant under re-parametrization of the orbit, so long as the parameters used are in one-to-one correspondence with $\{\en,\ang,\Q\}$. }, which provides yet another, independent, comparison point in the strong field, this time utilizing eccentric orbits. 

As a first example, one could consider the function $\Omega_{\varphi}(\Omega_r)$ along the singular curve in the parameter space of nonrotating binaries (Fig.\ \ref{fig:Schwarzschild_Omega_phi_e} shows this curve in the test-particle limit). In principle, one could compute this function in the SF approximation (i.e., order by order in the mass ratio), and perhaps also in fully nonlinear NR, making for an interesting comparison. There may be a way of using the results of such a calculation to calibrate the potentials of EOB theory in the strong field, although how this could be done in practice is yet unclear \cite{Damour-priv_comm}. Comparison with existing PN expressions could test the performance of the PN expansion in the strong field. A more constructive synergy could be achieved within the recent ``phenomenological'' approach to PN calculations, whereby high-order terms in the PN expansion are determined by fitting to numerical data from SF or NR calculations \cite{Blanchet_etal1,Blanchet_etal2}. A faithful phenomenological PN model would need to be able to recover the singular curve in the strong field, perhaps through the inclusion of suitable ``poles'' in PN expressions. 

Finally, let us mention the intriguing possibility that isofrequency pairing in astrophysical black holes (e.g., between clumps of accreting matter) could have observational implications. The question is worth asking because we are at an era where astronomical observations in a range of electromagnetic wavelengths routinely peer into processes deep in the strong-field potentials of accreting black holes. Quasi-periodic oscillations (QPOs) in x-rays from accreting black-hole systems probe the innermost regions of accretion disks \cite{Remillard-McClintock}, and (to a lesser extent) so do x-ray flares from the Galactic center \cite{Eckart_etal}. Could the peculiar strong-gravity phenomenon of isofrequency pairing have a dynamical effect on matter orbiting the black hole, perhaps through resonant interaction? Although admittedly far-fetched, this possibility deserves exploration.

\section*{Acknowledgements}

We thank Sam Dolan, Steve Drasco, Carsten Gundlach, Scott Hughes, Amos Ori and Eric Poisson for helpful discussions. We are also grateful to Maarten van de Meent for feedback on the first version of this paper. NWs work was supported by STFC through a studentship grant and by the Irish Research Council, which is funded under the National Development Plan for Ireland. LB acknowledges support from the European Research Council under grant No. 304978, and from STFC through grant number PP/E001025/1.


\begin{thebibliography}{31}%
\makeatletter
\providecommand \@ifxundefined [1]{%
 \@ifx{#1\undefined}
}%
\providecommand \@ifnum [1]{%
 \ifnum #1\expandafter \@firstoftwo
 \else \expandafter \@secondoftwo
 \fi
}%
\providecommand \@ifx [1]{%
 \ifx #1\expandafter \@firstoftwo
 \else \expandafter \@secondoftwo
 \fi
}%
\providecommand \natexlab [1]{#1}%
\providecommand \enquote  [1]{``#1''}%
\providecommand \bibnamefont  [1]{#1}%
\providecommand \bibfnamefont [1]{#1}%
\providecommand \citenamefont [1]{#1}%
\providecommand \href@noop [0]{\@secondoftwo}%
\providecommand \href [0]{\begingroup \@sanitize@url \@href}%
\providecommand \@href[1]{\@@startlink{#1}\@@href}%
\providecommand \@@href[1]{\endgroup#1\@@endlink}%
\providecommand \@sanitize@url [0]{\catcode `\\12\catcode `\$12\catcode
  `\&12\catcode `\#12\catcode `\^12\catcode `\_12\catcode `\%12\relax}%
\providecommand \@@startlink[1]{}%
\providecommand \@@endlink[0]{}%
\providecommand \url  [0]{\begingroup\@sanitize@url \@url }%
\providecommand \@url [1]{\endgroup\@href {#1}{\urlprefix }}%
\providecommand \urlprefix  [0]{URL }%
\providecommand \Eprint [0]{\href }%
\providecommand \doibase [0]{http://dx.doi.org/}%
\providecommand \selectlanguage [0]{\@gobble}%
\providecommand \bibinfo  [0]{\@secondoftwo}%
\providecommand \bibfield  [0]{\@secondoftwo}%
\providecommand \translation [1]{[#1]}%
\providecommand \BibitemOpen [0]{}%
\providecommand \bibitemStop [0]{}%
\providecommand \bibitemNoStop [0]{.\EOS\space}%
\providecommand \EOS [0]{\spacefactor3000\relax}%
\providecommand \BibitemShut  [1]{\csname bibitem#1\endcsname}%
\let\auto@bib@innerbib\@empty
%</preamble>
\bibitem [{\citenamefont {{Carter}}(1968)}]{Carter}%
  \BibitemOpen
  \bibfield  {author} {\bibinfo {author} {\bibfnamefont {B.}~\bibnamefont
  {{Carter}}},\ }\href {\doibase 10.1103/PhysRev.174.1559} {\bibfield
  {journal} {\bibinfo  {journal} {Physical Review}\ }\textbf {\bibinfo {volume}
  {174}},\ \bibinfo {pages} {1559} (\bibinfo {year} {1968})}\BibitemShut
  {NoStop}%
\bibitem [{\citenamefont {{Bardeen}}\ \emph {et~al.}(1972)\citenamefont
  {{Bardeen}}, \citenamefont {{Press}},\ and\ \citenamefont
  {{Teukolsky}}}]{Bardeen-Press-Teukolsky}%
  \BibitemOpen
  \bibfield  {author} {\bibinfo {author} {\bibfnamefont {J.~M.}\ \bibnamefont
  {{Bardeen}}}, \bibinfo {author} {\bibfnamefont {W.~H.}\ \bibnamefont
  {{Press}}}, \ and\ \bibinfo {author} {\bibfnamefont {S.~A.}\ \bibnamefont
  {{Teukolsky}}},\ }\href {\doibase 10.1086/151796} {\bibfield  {journal}
  {\bibinfo  {journal} {\apj}\ }\textbf {\bibinfo {volume} {178}},\ \bibinfo
  {pages} {347} (\bibinfo {year} {1972})}\BibitemShut {NoStop}%
\bibitem [{\citenamefont {Wilkins}(1972)}]{Wilkins}%
  \BibitemOpen
  \bibfield  {author} {\bibinfo {author} {\bibfnamefont {D.~C.}\ \bibnamefont
  {Wilkins}},\ }\href {\doibase 10.1103/PhysRevD.5.814} {\bibfield  {journal}
  {\bibinfo  {journal} {Phys. Rev. D}\ }\textbf {\bibinfo {volume} {5}},\
  \bibinfo {pages} {814} (\bibinfo {year} {1972})}\BibitemShut {NoStop}%
\bibitem [{\citenamefont {Chandrasekhar}(1992)}]{Chandra}%
  \BibitemOpen
  \bibfield  {author} {\bibinfo {author} {\bibfnamefont {S.}~\bibnamefont
  {Chandrasekhar}},\ }\href@noop {} {\emph {\bibinfo {title} {The Mathematical
  Theory of Black Holes}}}\ (\bibinfo  {publisher} {Oxford University Press},\
  \bibinfo {year} {1992})\BibitemShut {NoStop}%
\bibitem [{\citenamefont {{Bicak}}\ \emph {et~al.}(1993)\citenamefont
  {{Bicak}}, \citenamefont {{Semerak}},\ and\ \citenamefont
  {{Hadrava}}}]{Bicak_etal}%
  \BibitemOpen
  \bibfield  {author} {\bibinfo {author} {\bibfnamefont {J.}~\bibnamefont
  {{Bicak}}}, \bibinfo {author} {\bibfnamefont {O.}~\bibnamefont {{Semerak}}},
  \ and\ \bibinfo {author} {\bibfnamefont {P.}~\bibnamefont {{Hadrava}}},\
  }\href@noop {} {\bibfield  {journal} {\bibinfo  {journal} {\mnras}\ }\textbf
  {\bibinfo {volume} {263}},\ \bibinfo {pages} {545} (\bibinfo {year}
  {1993})}\BibitemShut {NoStop}%
\bibitem [{\citenamefont {{Schmidt}}(2002)}]{Schmidt}%
  \BibitemOpen
  \bibfield  {author} {\bibinfo {author} {\bibfnamefont {W.}~\bibnamefont
  {{Schmidt}}},\ }\href {\doibase 10.1088/0264-9381/19/10/314} {\bibfield
  {journal} {\bibinfo  {journal} {Classical and Quantum Gravity}\ }\textbf
  {\bibinfo {volume} {19}},\ \bibinfo {pages} {2743} (\bibinfo {year}
  {2002})}\BibitemShut {NoStop}%
\bibitem [{\citenamefont {{Drasco}}\ and\ \citenamefont
  {{Hughes}}(2004)}]{Drasco-Hughes}%
  \BibitemOpen
  \bibfield  {author} {\bibinfo {author} {\bibfnamefont {S.}~\bibnamefont
  {{Drasco}}}\ and\ \bibinfo {author} {\bibfnamefont {S.~A.}\ \bibnamefont
  {{Hughes}}},\ }\href {\doibase 10.1103/PhysRevD.69.044015} {\bibfield
  {journal} {\bibinfo  {journal} {\prd}\ }\textbf {\bibinfo {volume} {69}},\
  \bibinfo {eid} {044015} (\bibinfo {year} {2004})},\ \Eprint
  {http://arxiv.org/abs/arXiv:astro-ph/0308479} {arXiv:astro-ph/0308479}
  \BibitemShut {NoStop}%
\bibitem [{\citenamefont {{Levin}}\ and\ \citenamefont
  {{Perez-Giz}}(2008)}]{Levin-Periz-Giz:BH_periodic_table}%
  \BibitemOpen
  \bibfield  {author} {\bibinfo {author} {\bibfnamefont {J.}~\bibnamefont
  {{Levin}}}\ and\ \bibinfo {author} {\bibfnamefont {G.}~\bibnamefont
  {{Perez-Giz}}},\ }\href {\doibase 10.1103/PhysRevD.77.103005} {\bibfield
  {journal} {\bibinfo  {journal} {\prd}\ }\textbf {\bibinfo {volume} {77}},\
  \bibinfo {eid} {103005} (\bibinfo {year} {2008})},\ \Eprint
  {http://arxiv.org/abs/0802.0459} {arXiv:0802.0459 [gr-qc]} \BibitemShut
  {NoStop}%
\bibitem [{\citenamefont {{Grossman}}\ \emph {et~al.}(2012)\citenamefont
  {{Grossman}}, \citenamefont {{Levin}},\ and\ \citenamefont
  {{Perez-Giz}}}]{Grossman-Levin-Periz-Giz}%
  \BibitemOpen
  \bibfield  {author} {\bibinfo {author} {\bibfnamefont {R.}~\bibnamefont
  {{Grossman}}}, \bibinfo {author} {\bibfnamefont {J.}~\bibnamefont {{Levin}}},
  \ and\ \bibinfo {author} {\bibfnamefont {G.}~\bibnamefont {{Perez-Giz}}},\
  }\href {\doibase 10.1103/PhysRevD.85.023012} {\bibfield  {journal} {\bibinfo
  {journal} {\prd}\ }\textbf {\bibinfo {volume} {85}},\ \bibinfo {eid} {023012}
  (\bibinfo {year} {2012})},\ \Eprint {http://arxiv.org/abs/1105.5811}
  {arXiv:1105.5811 [gr-qc]} \BibitemShut {NoStop}%
\bibitem [{\citenamefont {{Fujita}}\ and\ \citenamefont
  {{Hikida}}(2009)}]{Fujita-Hikida}%
  \BibitemOpen
  \bibfield  {author} {\bibinfo {author} {\bibfnamefont {R.}~\bibnamefont
  {{Fujita}}}\ and\ \bibinfo {author} {\bibfnamefont {W.}~\bibnamefont
  {{Hikida}}},\ }\href {\doibase 10.1088/0264-9381/26/13/135002} {\bibfield
  {journal} {\bibinfo  {journal} {Classical and Quantum Gravity}\ }\textbf
  {\bibinfo {volume} {26}},\ \bibinfo {pages} {135002} (\bibinfo {year}
  {2009})}\BibitemShut {NoStop}%
\bibitem [{Note1()}]{Note1}%
  \BibitemOpen
  \bibinfo {note} {That the mapping $\protect \{r_{\protect \rm p},r_{\protect
  \rm a},\theta _{\protect \rm min}\protect \}\leftrightarrow \protect
  \{\protect \mathcal {E}, \protect \mathcal {L}_z, \protect \mathcal
  {Q}\protect \}$ is one-to-one can be establishing in the following way. We
  first note that Schmidt \cite {Schmidt} provides formula for $\protect
  \{\protect \mathcal {E}, \protect \mathcal {L}_z, \protect \mathcal
  {Q}\protect \}$ in terms of $(p,e,\theta _\protect \text {min})$, and that
  there is a bijection between $(p,e)\leftrightarrow \protect \{r_{\protect \rm
  p},r_{\protect \rm a}\protect \}$ (straightforward to see from Eqs.~\protect
  \textup {\hbox {\mathsurround \z@ \protect \normalfont (\ignorespaces \ref
  {pe}\unskip \@@italiccorr )}} and their inverse). Furthermore Eqs.~\protect
  \textup {\hbox {\mathsurround \z@ \protect \normalfont (\ignorespaces \ref
  {EOMr}\unskip \@@italiccorr )}} and \protect \textup {\hbox {\mathsurround
  \z@ \protect \normalfont (\ignorespaces \ref {EOMq}\unskip \@@italiccorr )}}
  imply that $r_1 \equiv r_a$, $r_2 \equiv r_p$ and $z_-$ (and hence $\theta
  _\protect \text {min}$) are given uniquely in terms of $\{\protect \mathcal
  {E},\protect \mathcal {L}_z,\protect \mathcal {Q}\}$. The existence of these
  relations asserts that the original mapping is one-to-one.}\BibitemShut
  {Stop}%
\bibitem [{\citenamefont {{Drasco}}(2009)}]{Drasco-2009}%
  \BibitemOpen
  \bibfield  {author} {\bibinfo {author} {\bibfnamefont {S.}~\bibnamefont
  {{Drasco}}},\ }\href {\doibase 10.1103/PhysRevD.79.104016} {\bibfield
  {journal} {\bibinfo  {journal} {\prd}\ }\textbf {\bibinfo {volume} {79}},\
  \bibinfo {eid} {104016} (\bibinfo {year} {2009})},\ \Eprint
  {http://arxiv.org/abs/0711.4644} {arXiv:0711.4644 [gr-qc]} \BibitemShut
  {NoStop}%
\bibitem [{\citenamefont {{Barack}}\ and\ \citenamefont
  {{Sago}}(2011)}]{Barack-Sago-2011}%
  \BibitemOpen
  \bibfield  {author} {\bibinfo {author} {\bibfnamefont {L.}~\bibnamefont
  {{Barack}}}\ and\ \bibinfo {author} {\bibfnamefont {N.}~\bibnamefont
  {{Sago}}},\ }\href {\doibase 10.1103/PhysRevD.83.084023} {\bibfield
  {journal} {\bibinfo  {journal} {\prd}\ }\textbf {\bibinfo {volume} {83}},\
  \bibinfo {eid} {084023} (\bibinfo {year} {2011})}\BibitemShut {NoStop}%
\bibitem [{\citenamefont {{Darwin}}(1961)}]{Darwin-1961}%
  \BibitemOpen
  \bibfield  {author} {\bibinfo {author} {\bibfnamefont {C.}~\bibnamefont
  {{Darwin}}},\ }\href {\doibase 10.1098/rspa.1961.0142} {\bibfield  {journal}
  {\bibinfo  {journal} {Royal Society of London Proceedings Series A}\ }\textbf
  {\bibinfo {volume} {263}},\ \bibinfo {pages} {39} (\bibinfo {year}
  {1961})}\BibitemShut {NoStop}%
\bibitem [{\citenamefont {{Cutler}}\ \emph {et~al.}(1994)\citenamefont
  {{Cutler}}, \citenamefont {{Kennefick}},\ and\ \citenamefont
  {{Poisson}}}]{Cutler-Kennefick-Poisson}%
  \BibitemOpen
  \bibfield  {author} {\bibinfo {author} {\bibfnamefont {C.}~\bibnamefont
  {{Cutler}}}, \bibinfo {author} {\bibfnamefont {D.}~\bibnamefont
  {{Kennefick}}}, \ and\ \bibinfo {author} {\bibfnamefont {E.}~\bibnamefont
  {{Poisson}}},\ }\href {\doibase 10.1103/PhysRevD.50.3816} {\bibfield
  {journal} {\bibinfo  {journal} {\prd}\ }\textbf {\bibinfo {volume} {50}},\
  \bibinfo {pages} {3816} (\bibinfo {year} {1994})}\BibitemShut {NoStop}%
\bibitem [{\citenamefont {{Glampedakis}}\ and\ \citenamefont
  {{Kennefick}}(2002)}]{Glampedakis-Kennefick}%
  \BibitemOpen
  \bibfield  {author} {\bibinfo {author} {\bibfnamefont {K.}~\bibnamefont
  {{Glampedakis}}}\ and\ \bibinfo {author} {\bibfnamefont {D.}~\bibnamefont
  {{Kennefick}}},\ }\href {\doibase 10.1103/PhysRevD.66.044002} {\bibfield
  {journal} {\bibinfo  {journal} {\prd}\ }\textbf {\bibinfo {volume} {66}},\
  \bibinfo {eid} {044002} (\bibinfo {year} {2002})}\BibitemShut {NoStop}%
\bibitem [{\citenamefont {{Levin}}\ and\ \citenamefont
  {{Perez-Giz}}(2009)}]{Levin-Perez-Giz}%
  \BibitemOpen
  \bibfield  {author} {\bibinfo {author} {\bibfnamefont {J.}~\bibnamefont
  {{Levin}}}\ and\ \bibinfo {author} {\bibfnamefont {G.}~\bibnamefont
  {{Perez-Giz}}},\ }\href {\doibase 10.1103/PhysRevD.79.124013} {\bibfield
  {journal} {\bibinfo  {journal} {\prd}\ }\textbf {\bibinfo {volume} {79}},\
  \bibinfo {eid} {124013} (\bibinfo {year} {2009})}\BibitemShut {NoStop}%
\bibitem [{\citenamefont {{Mino}}(2003)}]{Mino}%
  \BibitemOpen
  \bibfield  {author} {\bibinfo {author} {\bibfnamefont {Y.}~\bibnamefont
  {{Mino}}},\ }\href {\doibase 10.1103/PhysRevD.67.084027} {\bibfield
  {journal} {\bibinfo  {journal} {\prd}\ }\textbf {\bibinfo {volume} {67}},\
  \bibinfo {eid} {084027} (\bibinfo {year} {2003})}\BibitemShut {NoStop}%
\bibitem [{\citenamefont {{Drasco}}\ and\ \citenamefont
  {{Hughes}}(2006)}]{Drasco-Hughes-2006}%
  \BibitemOpen
  \bibfield  {author} {\bibinfo {author} {\bibfnamefont {S.}~\bibnamefont
  {{Drasco}}}\ and\ \bibinfo {author} {\bibfnamefont {S.~A.}\ \bibnamefont
  {{Hughes}}},\ }\href {\doibase 10.1103/PhysRevD.73.024027} {\bibfield
  {journal} {\bibinfo  {journal} {\prd}\ }\textbf {\bibinfo {volume} {73}},\
  \bibinfo {eid} {024027} (\bibinfo {year} {2006})},\ \Eprint
  {http://arxiv.org/abs/arXiv:gr-qc/0509101} {arXiv:gr-qc/0509101} \BibitemShut
  {NoStop}%
\bibitem [{\citenamefont {{Sundararajan}}(2008)}]{Sundararajan}%
  \BibitemOpen
  \bibfield  {author} {\bibinfo {author} {\bibfnamefont {P.~A.}\ \bibnamefont
  {{Sundararajan}}},\ }\href {\doibase 10.1103/PhysRevD.77.124050} {\bibfield
  {journal} {\bibinfo  {journal} {\prd}\ }\textbf {\bibinfo {volume} {77}},\
  \bibinfo {eid} {124050} (\bibinfo {year} {2008})}\BibitemShut {NoStop}%
\bibitem [{\citenamefont {{Le Tiec}}\ \emph {et~al.}(2011)\citenamefont {{Le
  Tiec}}, \citenamefont {{Mrou{\'e}}}, \citenamefont {{Barack}}, \citenamefont
  {{Buonanno}}, \citenamefont {{Pfeiffer}}, \citenamefont {{Sago}},\ and\
  \citenamefont {{Taracchini}}}]{LeTiec_etal-2011}%
  \BibitemOpen
  \bibfield  {author} {\bibinfo {author} {\bibfnamefont {A.}~\bibnamefont {{Le
  Tiec}}}, \bibinfo {author} {\bibfnamefont {A.~H.}\ \bibnamefont
  {{Mrou{\'e}}}}, \bibinfo {author} {\bibfnamefont {L.}~\bibnamefont
  {{Barack}}}, \bibinfo {author} {\bibfnamefont {A.}~\bibnamefont
  {{Buonanno}}}, \bibinfo {author} {\bibfnamefont {H.~P.}\ \bibnamefont
  {{Pfeiffer}}}, \bibinfo {author} {\bibfnamefont {N.}~\bibnamefont {{Sago}}},
  \ and\ \bibinfo {author} {\bibfnamefont {A.}~\bibnamefont {{Taracchini}}},\
  }\href {\doibase 10.1103/PhysRevLett.107.141101} {\bibfield  {journal}
  {\bibinfo  {journal} {Physical Review Letters}\ }\textbf {\bibinfo {volume}
  {107}},\ \bibinfo {eid} {141101} (\bibinfo {year} {2011})},\ \Eprint
  {http://arxiv.org/abs/1106.3278} {arXiv:1106.3278 [gr-qc]} \BibitemShut
  {NoStop}%
\bibitem [{\citenamefont {{Barack}}\ and\ \citenamefont
  {{Sago}}(2009)}]{Barack-Sago-PRL}%
  \BibitemOpen
  \bibfield  {author} {\bibinfo {author} {\bibfnamefont {L.}~\bibnamefont
  {{Barack}}}\ and\ \bibinfo {author} {\bibfnamefont {N.}~\bibnamefont
  {{Sago}}},\ }\href {\doibase 10.1103/PhysRevLett.102.191101} {\bibfield
  {journal} {\bibinfo  {journal} {Physical Review Letters}\ }\textbf {\bibinfo
  {volume} {102}},\ \bibinfo {eid} {191101} (\bibinfo {year} {2009})},\ \Eprint
  {http://arxiv.org/abs/0902.0573} {arXiv:0902.0573 [gr-qc]} \BibitemShut
  {NoStop}%
\bibitem [{\citenamefont {{Damour}}(2010)}]{Damour-2010}%
  \BibitemOpen
  \bibfield  {author} {\bibinfo {author} {\bibfnamefont {T.}~\bibnamefont
  {{Damour}}},\ }\href {\doibase 10.1103/PhysRevD.81.024017} {\bibfield
  {journal} {\bibinfo  {journal} {\prd}\ }\textbf {\bibinfo {volume} {81}},\
  \bibinfo {eid} {024017} (\bibinfo {year} {2010})},\ \Eprint
  {http://arxiv.org/abs/0910.5533} {arXiv:0910.5533 [gr-qc]} \BibitemShut
  {NoStop}%
\bibitem [{\citenamefont {{Favata}}(2011)}]{Favata-2011}%
  \BibitemOpen
  \bibfield  {author} {\bibinfo {author} {\bibfnamefont {M.}~\bibnamefont
  {{Favata}}},\ }\href {\doibase 10.1103/PhysRevD.83.024028} {\bibfield
  {journal} {\bibinfo  {journal} {\prd}\ }\textbf {\bibinfo {volume} {83}},\
  \bibinfo {eid} {024028} (\bibinfo {year} {2011})},\ \Eprint
  {http://arxiv.org/abs/1010.2553} {arXiv:1010.2553 [gr-qc]} \BibitemShut
  {NoStop}%
\bibitem [{\citenamefont {{Barack}}\ \emph {et~al.}(2010)\citenamefont
  {{Barack}}, \citenamefont {{Damour}},\ and\ \citenamefont
  {{Sago}}}]{Barack-Damour-Sago}%
  \BibitemOpen
  \bibfield  {author} {\bibinfo {author} {\bibfnamefont {L.}~\bibnamefont
  {{Barack}}}, \bibinfo {author} {\bibfnamefont {T.}~\bibnamefont {{Damour}}},
  \ and\ \bibinfo {author} {\bibfnamefont {N.}~\bibnamefont {{Sago}}},\ }\href
  {\doibase 10.1103/PhysRevD.82.084036} {\bibfield  {journal} {\bibinfo
  {journal} {\prd}\ }\textbf {\bibinfo {volume} {82}},\ \bibinfo {eid} {084036}
  (\bibinfo {year} {2010})},\ \Eprint {http://arxiv.org/abs/1008.0935}
  {arXiv:1008.0935 [gr-qc]} \BibitemShut {NoStop}%
\bibitem [{Note2()}]{Note2}%
  \BibitemOpen
  \bibinfo {note} {By ``invariant structure'' we refer to the fact that the
  singular surface in the frequency space is invariant under re-parametrization
  of the orbit, so long as the parameters used are in one-to-one correspondence
  with $\protect \{\protect \mathcal {E},\protect \mathcal {L}_z,\protect
  \mathcal {Q}\protect \}$.}\BibitemShut {Stop}%
\bibitem [{\citenamefont {{Damour}}()}]{Damour-priv_comm}%
  \BibitemOpen
  \bibfield  {author} {\bibinfo {author} {\bibfnamefont {T.}~\bibnamefont
  {{Damour}}},\ }\href@noop {} {}\bibinfo {note} {Private
  communication}\BibitemShut {NoStop}%
\bibitem [{\citenamefont {{Blanchet}}\ \emph {et~al.}(2010)\citenamefont
  {{Blanchet}}, \citenamefont {{Detweiler}}, \citenamefont {{Le Tiec}},\ and\
  \citenamefont {{Whiting}}}]{Blanchet_etal1}%
  \BibitemOpen
  \bibfield  {author} {\bibinfo {author} {\bibfnamefont {L.}~\bibnamefont
  {{Blanchet}}}, \bibinfo {author} {\bibfnamefont {S.}~\bibnamefont
  {{Detweiler}}}, \bibinfo {author} {\bibfnamefont {A.}~\bibnamefont {{Le
  Tiec}}}, \ and\ \bibinfo {author} {\bibfnamefont {B.~F.}\ \bibnamefont
  {{Whiting}}},\ }\href {\doibase 10.1103/PhysRevD.81.084033} {\bibfield
  {journal} {\bibinfo  {journal} {\prd}\ }\textbf {\bibinfo {volume} {81}},\
  \bibinfo {eid} {084033} (\bibinfo {year} {2010})},\ \Eprint
  {http://arxiv.org/abs/1002.0726} {arXiv:1002.0726 [gr-qc]} \BibitemShut
  {NoStop}%
\bibitem [{\citenamefont {{Blanchet}}\ \emph {et~al.}(2011)\citenamefont
  {{Blanchet}}, \citenamefont {{Detweiler}}, \citenamefont {{Le Tiec}},\ and\
  \citenamefont {{Whiting}}}]{Blanchet_etal2}%
  \BibitemOpen
  \bibfield  {author} {\bibinfo {author} {\bibfnamefont {L.}~\bibnamefont
  {{Blanchet}}}, \bibinfo {author} {\bibfnamefont {S.}~\bibnamefont
  {{Detweiler}}}, \bibinfo {author} {\bibfnamefont {A.}~\bibnamefont {{Le
  Tiec}}}, \ and\ \bibinfo {author} {\bibfnamefont {B.~F.}\ \bibnamefont
  {{Whiting}}},\ }\enquote {\bibinfo {title} {{High-Accuracy Comparison Between
  the Post-Newtonian and Self-Force Dynamics of Black-Hole Binaries}},}\ in\
  \href {\doibase 10.1007/978-90-481-3015-3_15} {\emph {\bibinfo {booktitle}
  {Mass and Motion in General Relativity}}},\ \bibinfo {editor} {edited by\
  \bibinfo {editor} {\bibfnamefont {L.}~\bibnamefont {{Blanchet}}}, \bibinfo
  {editor} {\bibfnamefont {A.}~\bibnamefont {{Spallicci}}}, \ and\ \bibinfo
  {editor} {\bibfnamefont {B.}~\bibnamefont {{Whiting}}}}\ (\bibinfo {year}
  {2011})\ pp.\ \bibinfo {pages} {415--442}\BibitemShut {NoStop}%
\bibitem [{\citenamefont {{Remillard}}\ and\ \citenamefont
  {{McClintock}}(2006)}]{Remillard-McClintock}%
  \BibitemOpen
  \bibfield  {author} {\bibinfo {author} {\bibfnamefont {R.~A.}\ \bibnamefont
  {{Remillard}}}\ and\ \bibinfo {author} {\bibfnamefont {J.~E.}\ \bibnamefont
  {{McClintock}}},\ }\href {\doibase 10.1146/annurev.astro.44.051905.092532}
  {\bibfield  {journal} {\bibinfo  {journal} {\araa}\ }\textbf {\bibinfo
  {volume} {44}},\ \bibinfo {pages} {49} (\bibinfo {year} {2006})},\ \Eprint
  {http://arxiv.org/abs/arXiv:astro-ph/0606352} {arXiv:astro-ph/0606352}
  \BibitemShut {NoStop}%
\bibitem [{\citenamefont {{Eckart}}\ \emph {et~al.}(2012)\citenamefont
  {{Eckart}}, \citenamefont {{Garc{\'{\i}}a-Mar{\'{\i}}n}}, \citenamefont
  {{Vogel}}, \citenamefont {{Teuben}}, \citenamefont {{Morris}}, \citenamefont
  {{Baganoff}}, \citenamefont {{Dexter}}, \citenamefont {{Sch{\"o}del}},
  \citenamefont {{Witzel}}, \citenamefont {{Valencia-S}}, \citenamefont
  {{Karas}}, \citenamefont {{Kunneriath}}, \citenamefont {{Bremer}},
  \citenamefont {{Straubmeier}}, \citenamefont {{Moser}}, \citenamefont
  {{Sabha}}, \citenamefont {{Buchholz}}, \citenamefont {{Zamaninasab}},
  \citenamefont {{Mu{\v z}i{\'c}}}, \citenamefont {{Moultaka}},\ and\
  \citenamefont {{Zensus}}}]{Eckart_etal}%
  \BibitemOpen
  \bibfield  {author} {\bibinfo {author} {\bibfnamefont {A.}~\bibnamefont
  {{Eckart}}}, \bibinfo {author} {\bibfnamefont {M.}~\bibnamefont
  {{Garc{\'{\i}}a-Mar{\'{\i}}n}}}, \bibinfo {author} {\bibfnamefont {S.~N.}\
  \bibnamefont {{Vogel}}}, \bibinfo {author} {\bibfnamefont {P.}~\bibnamefont
  {{Teuben}}}, \bibinfo {author} {\bibfnamefont {M.~R.}\ \bibnamefont
  {{Morris}}}, \bibinfo {author} {\bibfnamefont {F.}~\bibnamefont
  {{Baganoff}}}, \bibinfo {author} {\bibfnamefont {J.}~\bibnamefont
  {{Dexter}}}, \bibinfo {author} {\bibfnamefont {R.}~\bibnamefont
  {{Sch{\"o}del}}}, \bibinfo {author} {\bibfnamefont {G.}~\bibnamefont
  {{Witzel}}}, \bibinfo {author} {\bibfnamefont {M.}~\bibnamefont
  {{Valencia-S}}}, \bibinfo {author} {\bibfnamefont {V.}~\bibnamefont
  {{Karas}}}, \bibinfo {author} {\bibfnamefont {D.}~\bibnamefont
  {{Kunneriath}}}, \bibinfo {author} {\bibfnamefont {M.}~\bibnamefont
  {{Bremer}}}, \bibinfo {author} {\bibfnamefont {C.}~\bibnamefont
  {{Straubmeier}}}, \bibinfo {author} {\bibfnamefont {L.}~\bibnamefont
  {{Moser}}}, \bibinfo {author} {\bibfnamefont {N.}~\bibnamefont {{Sabha}}},
  \bibinfo {author} {\bibfnamefont {R.}~\bibnamefont {{Buchholz}}}, \bibinfo
  {author} {\bibfnamefont {M.}~\bibnamefont {{Zamaninasab}}}, \bibinfo {author}
  {\bibfnamefont {K.}~\bibnamefont {{Mu{\v z}i{\'c}}}}, \bibinfo {author}
  {\bibfnamefont {J.}~\bibnamefont {{Moultaka}}}, \ and\ \bibinfo {author}
  {\bibfnamefont {J.~A.}\ \bibnamefont {{Zensus}}},\ }\href {\doibase
  10.1088/1742-6596/372/1/012022} {\bibfield  {journal} {\bibinfo  {journal}
  {Journal of Physics Conference Series}\ }\textbf {\bibinfo {volume} {372}},\
  \bibinfo {pages} {012022} (\bibinfo {year} {2012})},\ \Eprint
  {http://arxiv.org/abs/1208.1135} {arXiv:1208.1135 [astro-ph.IM]} \BibitemShut
  {NoStop}%
\end{thebibliography}
\end{document}